\documentclass[aps,notitlepage,twocolumn,prb]{revtex4-1}
\usepackage{amsfonts}
\usepackage{amsmath}
\usepackage{amssymb}
\usepackage[final,dvips]{graphicx}
\providecommand{\U}[1]{\protect\rule{.1in}{.1in}}
\begin{document}
\title{Phase diagram of Josephson junction between $s$ and $s_{\pm}$
superconductors in dirty limit}
\author{A. E. Koshelev}
\affiliation{Materials Science Division, Argonne National Laboratory, Argonne,
Illinois 60439}
\date{\today }

\begin{abstract}
The $s_{\pm}$ state in which the order parameter has different signs in different
bands is a leading candidate for the superconducting state in the iron-based
superconductors. We investigate a Josephson junction between $s$ and $s_{\pm}$
superconductors within microscopic theory. Frustration, caused by interaction of the
$s$-wave gap parameter with the opposite-sign gaps of the $s_{\pm}$ superconductor,
leads to nontrivial phase diagram. When the partial Josephson coupling energy
between the $s$-wave superconductor and one of the $s_{\pm}$ bands dominates, s-wave
gap parameter aligns with the order parameter in this band. In this case the partial
Josephson energies have different signs corresponding to signs of the gap
parameters. In the case of strong frustration, corresponding to almost complete
compensation of the total Josephson energy, a nontrivial time-reversal-symmetry
breaking (TRSB) state realizes. In this state all gap parameters become essentially
complex. As a consequence, this state provides realization for so-called
$\phi$-junction with finite phase difference in the ground state. The width of the
TRSB state region is determined by the second harmonic in Josephson current,
$\propto\sin(2\phi)$, which appears in the second order with respect to the boundary
transparency. Using the microscopic theory, we establish a range of parameters where
different states are realized. Our analysis shows insufficiency of the simple
phenomenological approach for treatment of this problem.
\end{abstract}
\maketitle

\section{Introduction}

The discovery of superconducting iron pnictides and chalcogenides is one of the most
remarkable recent achievements in the condensed-matter physics. A rapid progress in
characterization of these materials and development of theoretical understanding has
been reflected in several reviews\cite{Reviews,ReviewsTheory}. The key feature of
these semimetallic materials is the multiple-band structure, the Fermi surface is
composed of several electron and hole pockets located near different points of the
Brillouin zone.

Superconductivity in the iron-based materials is likely to be unconventional. There
is a theoretical consensus that the electron-phonon interaction is not strong enough
to explain high transition temperatures.\cite{BoeriPRL08}  In several theoretical
papers it was suggested that superconductivity is mediated by spin fluctuations
leading to an unusual superconducting state in which the order parameter has
opposite signs in the electron and hole bands ($s_{\pm}$
state).\cite{MazinPRL08,KurokiPRL08,SeoPRL08,GraserNJP09,CvetkovicEPL09}
Experimental verification of this theoretical proposal became one of the major
challenges in the field. Probing the relative sign of the order parameter in
different bands is not trivial and the structure of superconducting state has not
been unambiguously established yet by experiment, even though several properties
consistent with the $s_{\pm}$ state have been revealed. An extensive critical review
of experiments both in favor and against the realization of the $s_{\pm}$ state in
iron-based superconductors has been done recently in Refs.\
\onlinecite{ReviewsTheory}. Shortly, the main experiments supporting the $s_{\pm}$
state include
\begin{itemize}
  \item Observation by the inelastic neutron scattering of the resonant magnetic mode below the
superconducting transition temperature.\cite{MagRes} Such a mode is expected for
the superconductors with the sign-changing order parameter. This mode was
observed in almost all compounds and its frequency scales approximately
proportional to the transition temperature.
  \item Microscopic coexistence of antiferromagnetism and superconductivity demonstrated in some
compounds within a narrow doping range, most clearly in
Ba[Fe$_{1-x}$Co$_{x}$]$_{2}$As$_{2}$.\cite{LaplacePRB09,Fernandes10} For the
case of the conventional $s_{++}$ state in which the order parameter has the
same sign in all bands, the spin-density wave (SDW) has a strong pair-breaking
effect on the bands connected with the SDW ordering wave vector. Such direct
pair breaking is absent if the order parameter in such bands has opposite signs
meaning that the SDW is much more compatible with the $s_{\pm}$ state than with
$s_{++}$ one.\cite{CoexTheory}
  \item The magnetic field dependence of the quasiparticle interference peaks studied by
  scanning tunneling spectroscopy in FeSe$_x$Te$_{1-x}$.\cite{HanaguriSci10}
\end{itemize}
On the other hand, discovery of the iron selenide compounds \emph{without hole band}
and with rather high transition temperatures, up to 30K \cite{K122}, questioned
universality of the $s_{\pm}$ state for all iron-based superconductors. Also, it
occurs that the iron-based superconductors are quite stable with respect to
disorder. As for the $s_{\pm}$ state the interband scattering is pair breaking,
stability with respect to disorder is frequently used as an argument against this
state. Therefore the structure of the order parameter in the iron-based
superconductors is an unresolved issue.

\begin{figure}[h]
\includegraphics[width=0.48\textwidth]{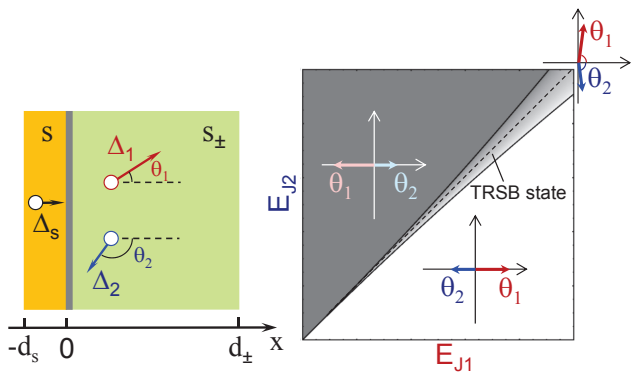}
\caption{\emph{Left part:}
A Josephson junction between $s$ and $s_{\pm}$ superconductors with the
gap parameters for a general complex state. \emph{Right part:} Generic phase
diagram of such Josephson junction. }
\label{SchemFig}
\end{figure}
One of the ways to probe unconventional superconductivity is to study Josephson
junctions and proximity effects with conventional superconductors.  In the case of
contact between s-wave and $s_{\pm}$ superconductors, frustration, caused by
interaction of the $s$-wave gap parameter with the opposite-sign gaps of $s_{\pm}$
superconductor leads to several anomalous features which were recognized and studied
in several theoretical
papers.\cite{Ng,Linder,SperstadPRB09,VS,Ota,Berg,ProxFingeprintEPL11,Lin,Levchenko}
For example, proximity with $s_{\pm}$ superconductor induces corrections to the
density of states of s-wave superconductor which, in principle, allow to identify
the signs of the order parameter in different bands of $s_{\pm}$
superconductor.\cite{ProxFingeprintEPL11} Particularly interesting is a possibility
of a time-reversal symmetry breaking (TRSB) state\cite{Ng,VS,Berg,Lin} in the
parameter range where the partial Josephson coupling energies between
$s$-superconductor and different $s_{\pm}$ bands almost exactly compensate each
other. Existing experiments on Josephson junctions between iron-based and
conventional superconductors have been reviewed in Ref.\ \onlinecite{SeidelSUST11}.
No anomalous features, however, have been reported so far.

For experimental realization of the TRSB state it is important to establish range of
parameters where such state can be expected. Up to now this state was studied using
mostly phenomenological models which are not rigorously justified. The purpose of
this paper is to develop microscopic description of transition between the aligned
and TRSB states. The paper is organized as follows. In Sec.\
\ref{Sec:TransSecondHarm} we present general consideration of the transition between
the aligned and TRSB state in the region where the partial Josephson energies almost
compensate each other. It is known that the width of the TRSB region is determined
by the second harmonic of the Josephson current, which in our situation appears in
the second order with respect to the coupling between superconductors.  In Sec.\
\ref{Sec:Ueq} we present the microscopic equations and boundary conditions
describing the contact between $s$-wave and $s_{\pm}$ superconductors in dirty limit
considered in this paper.  In Sec.\ \ref{Sec:Corrections} we consider corrections to
the Green's functions and gap parameters induced by the interface. Computation
details of these corrections are presented in Appendix \ref{App-WeakCoupl}.  The
proximity-induced corrections to the Green's function determine the second harmonic
in the Josephson current, which is considered in Sec.\ \ref{Sec:SecondHarm}. We
present both general formulas for different contributions to the second harmonic and
simple analytical results for the most relevant limiting cases. We also reveal the
dominating contribution to the second harmonic. Using these results we analyze in
Sec.\ \ref{Sec:TRSBWidth} the width of the TRSB region and its shrinking with
increasing temperature. Finally, in Sec.\ \ref{Sec:DoSTRSB} we consider
proximity-induced corrections to the density of states of the $s$-wave
superconductor within the TRSB region.

\section{General consideration of the transition region in the weak-coupling
case \label{Sec:TransSecondHarm}}

We consider the Josephson junction between $s$-wave and two-band $s_{\pm}$
superconductors, see Fig.\ \ref{SchemFig} (left). In the weak-coupling limit this
system is characterized by the partial Josephson coupling energies between
$s$-superconductor and $s_{\pm}$ bands, $E_{J\alpha}$ with $\alpha$ being the band
index. Typically, the $s$-wave gap parameter aligns along the  $s_{\pm}$ gap with
which it has larger coupling energy. In this aligned state the partial Josephson
coupling energies are positive and negative for the aligned and anti-aligned bands
correspondingly. Nontrivial behavior is expected in the case of strong frustration
when the total Josephson energy is close to zero. This happens when absolute values
of the Josephson energies for the opposite-sign bands are close, $|E_{J1}|\approx
|E_{J2}|$. Phenomenologically, the phase diagram can be described by the model of
the frustrated Josephson junction considered in several papers \cite{Ng} which
provides correct qualitative description. However, in general, this model does not
describe the system quantitatively, because, ignoring the fermionic degrees of
freedom in the $s_{\pm}$ superconductor, it does not treat correctly its interband
energy. For the weak-coupling regime, however, the transitional region between the
two aligned states can be treated following the same reasoning as for the transition
between $0$ and $\pi$ junctions, see, e.g., Refs.
\onlinecite{BuzdinReview,SperstadPRB09}. In the vicinity of transition the linear
approximation for the coupling between the superconductors becomes insufficient and
the total Josephson energy can be represented as
\begin{equation}
\mathcal{E(\phi)}=(E_{J1}-E_{J2})(1-\cos\phi)+\frac{E_{J}^{(2)}}{2}
(1-\cos2\phi),
\end{equation}
where $\phi$ is the phase difference between $\Delta_{s}$ and $\Delta_{1}$ and the
term $E_{J}^{(2)}$ appears in the second order with respect to the boundary
transparency. This corresponds to the Josephson current
\begin{equation}
j\mathcal{(\phi)}=(j_{J1}-j_{J2})\sin\phi+j_{J}^{(2)}\sin2\phi
\end{equation}
with $j_{J\alpha}=(2\pi c/\Phi_0)E_{J\alpha}$. The intermediate TRSB state exists
only if the sign of the second-harmonic is negative $E_{J}^{(2)},j_{J}^{(2)}<0$. In
this case in the region $|j_{J1}-j_{J2}|<2j_{J}^{(2)}$ the ground-state phase
difference is given by
\begin{equation}
\cos\phi_{0}=(j_{J1}-j_{J2})/(2|j_{J}^{(2)}|).
\end{equation}
It smoothly transforms between $0$ and $\pi$ when the difference $j_{J1}-j_{J2}$
changes from $2|j_{J}^{(2)}|$ to $-2|j_{J}^{(2)}|$. Therefore the TRSB state also
provides realizations of so-called $\phi$-junction \cite{fi-junction} in which a
finite phase difference exists in ground state leading to several anomalous
properties. In the case $E_{J}^{(2)},j_{J}^{(2)}>0$ the transition between the two
aligned states is of the first order and the TRSB state is not realized.

The simplest phenomenological description is the frustrated Josephson junction model
in which the tilt of the relative phase between two gap parameters of $s_{\pm}$
superconductors is described by the energy $\mathcal{E}_{12}
\cos(\theta_{1}-\theta_{2})$. In this model the amplitude of the second harmonic is
given by
\[
j_{J}^{(2)}=-\bar{j}_{J}\bar{E}_{J}/(2\mathcal{E}_{12})
\]
with $\bar{j}_{J}=(j_{J1}+j_{J2})/2$ and $\bar{E}_{J}=(E_{J1}+E_{J2})/2$. Our
further microscopic analysis shows that this result is only valid for a special
situation of very weak coupling between the bands of the $s_{\pm}$ superconductor.
We will compute the second harmonic in general case within microscopic approach.

\section{Equations and boundary conditions}
\label{Sec:Ueq}

In this section we write down equations and boundary conditions for the simple
microscopic model describing a ``sandwich'', consisting of slabs of two-band
$s_{\pm}$ superconductor with thickness $d_{\mathrm{\pm}}$ and a single-band
$s$-wave superconductor with thickness $d_{s}$, as shown on Fig.\ \ref{SchemFig}. We
denote the bulk critical temperatures of the $s$-wave and $s_{\pm}$ superconductors
as $T_{c}^{s}$ and $T_{c}$, respectively. The $x\!=\!0$ plane coincides with the
interface between the superconductors. The main assumption of our description is
that both superconductors are in dirty limit but the interband scattering in the
$s_{\pm}$ superconductor is negligible. In this case bulk superconductivity is
described by quasiclassical Usadel equations \cite{Usadel} with boundary conditions
derived in Ref.\ \onlinecite{KL}. The conventional proximity effects were
extensively explored within this approach in Ref.\ \onlinecite{Golubov1}. This
description was later generalized to conventional two-band superconductors in Ref.\
\onlinecite{Golubov2}. This model have been already used to describe some anomalous
properties of the $s$/$s_{\pm}$ interface in Refs.\
\onlinecite{ProxFingeprintEPL11}.

Both superconductors are described by the gap parameters, $\Delta(x)$, and the
impurity averaged Green's functions, which have regular and anomalous components,
$G(x,\omega)$ and $F(x,\omega)$, with $G^2+|F|^2=1$, where $\omega=2 \pi T (n+1/2)$
are the Matsubara frequencies. In the following, we will use subscript ``s'' for the
s-wave superconductor and subscript ``$\alpha$'' for the $\alpha$-band of the
$s_{\pm}$ superconductor and skip subscripts in relations applicable for both
superconductors. Further, we employ so-called
$\Phi$-parametrization\cite{Golubov1,Golubov2} in which the function $\Phi=\omega
F/G$ is used instead of $F$. In this case $G=\omega/\sqrt{\omega^{2}+|\Phi|^{2}}$.
For the $s$-wave superconductor ($-d_{s}<x<0$), the equations for the  Green's
functions $G_{s}$ and $\Phi_{s}$ and the self-consistency equation are:
\begin{subequations}
\begin{align}
&  \frac{D_{s}}{2\omega G_{s}}\frac{d}{dx}\left[  G_{s}^{2}\frac{d\Phi_{s}}
{dx}\right]  -\Phi_{s} =-\Delta_{s},\label{Ueqs1}\\
&2\pi T\sum_{\omega >0}\left(
\frac{\Phi _{s}}{\sqrt{\omega ^{2}+|\Phi _{s}|^{2}}}-\frac{\Delta _{s}}{\omega }\right)
+\Delta _{s}\ln \frac{T_{c}^{s}}{T}=0
\label{Ueqs2}
\end{align}
\end{subequations}
Correspondingly, for the $s_{\pm}$-superconductor, $0<x<d_{\mathrm{\pm}}$ we have
\begin{subequations}
\begin{align}
&\frac{D_{\alpha}}{2\omega G_{\alpha}}\frac{d}{dx}\left[  G_{\alpha}^{2}
\frac{d\Phi_{\alpha}}{dx}\right]  -\Phi_{\alpha}   =\!-\Delta_{\alpha}
,\label{Ueqspm1}\\
&2\pi T\sum_{\beta,\omega>0}\!\lambda_{\alpha\beta}
\frac{\Phi_{ \beta} }{\sqrt{\omega ^{2}+|\Phi _{\beta }|^{2}}}=\Delta_{\alpha}.
\label{Ueqspm2}
\end{align}
where $\lambda_{\alpha \beta}$ is the coupling-constants matrix and $\alpha$,
$\beta$ are the band indices. In Eq.\ (\ref{Ueqspm1}) we neglected the interband
impurity scattering. For the case of $s_{\pm}$ superconductor we consider here
$\Delta_{1} \Delta_{2}<0$. This is realized when $\lambda_{12},\lambda_{21}<0$. The
diffusion coefficients $D_{\{s, \alpha\}}$ are related to the conductivities
$\sigma_{\{s, \alpha\}}$ as $\sigma_{\{s, \alpha\}}=e^{2}\nu_{\{s, \alpha\}} D_{\{s,
\alpha\}}$, where $\nu_{\{s, \alpha\}}$ are the normal densities of states (DoS).
The ratio of the off-diagonal coupling constants is given by the ratio of partial
normal DoSs, $\lambda_{\alpha\beta}/\lambda_{\beta\alpha}=\nu_{\beta}/\nu_{\alpha}$.
It is convenient to normalize all energy parameters ($\omega$ and gaps on
\emph{both} sides) to the same scale $\pi T_{c}$. We also introduce the coherence
lengths $\xi_{\alpha}= \sqrt{D_{\alpha}/2 \pi T_{c}}$ and $\xi^{\ast} _{s}=\sqrt{
D_{s}/2 \pi T_{c}}$ (note that $\xi^{\ast}_{s}$ is related to the bulk coherence
length of the $s$-wave superconductor by $\xi_{s}=\xi^{\ast }_{s}
\sqrt{T_{c}/T^{s}_{c}}$ ).

The bulk equations have to be supplemented with the boundary conditions at the
interface separating two superconductors. These conditions relate the Green's
functions and their derivatives at the interface and can be written as
\cite{KL,Golubov2}
\end{subequations}
\begin{subequations}
\begin{align}
&  \xi^{\ast}_{s}G_{s} \frac{d\Phi_{s}}{dx} = \sum_{\alpha}\frac{G_{\alpha}
}{\tilde{\gamma}_{B \alpha}}(\Phi_{\alpha}-\Phi_{s}),\label{BCgen1}\\
&  \xi_{\alpha} G_{\alpha}\frac{d\Phi_{\alpha}}{dx} = -\frac{G_{s}}{\gamma_{B
\alpha} } (\Phi_{s}-\Phi_{\alpha}), \label{BCgen2}
\end{align}
\end{subequations}
for $x=0$, where $\alpha$ is the band index. Here the coupling parameters, $\tilde
{\gamma}_{B\alpha}$ and $\tilde{\gamma}_{B\alpha}$ are proportional to the partial
boundary resistances $R_{B \alpha}$,
\begin{equation}
\tilde{\gamma}_{B\alpha}=\frac{R_{B \alpha}} {\rho_{s}\xi_{s}^{\ast}},\ \ \
\gamma_{B \alpha}=\frac{R_{B \alpha}} {\rho_{\alpha}\xi_{\alpha}},
\label{gBa}
\end{equation}
where $\rho_{\{s,\alpha\}}\!=\!1/\sigma_{\{s,\alpha\}}$ are the bulk resistivities.
We will also use the ratios of these parameters
\begin{equation}
\gamma_{\alpha}=\frac{\tilde{\gamma}_{B\alpha}}{\gamma_{B \alpha}}=\frac
{\rho_{\alpha}\xi_{\alpha}}{\rho_{s}\xi^{\ast}_{s}},
\label{ga}
\end{equation}
that are bulk parameters characterizing the relative ``metalicity'' of the s-wave
superconductor and $\alpha$ band. In particular, large $\gamma_{\alpha}$ implies
that the $s$-wave material is more metallic than the $\alpha$ band on the $s_{\pm}$
side. The parameters $\gamma_{\alpha}$, $\lambda_{\alpha\beta}$, and $\xi_{\alpha}$
are not fully independent, the ratio of $\gamma _{\alpha}$ obeys the following
relation
\begin{equation}
\frac{\gamma_{1}}{\gamma_{2}}=\frac{\nu_{2}\xi_{2}}{\nu_{1}\xi_{1}}=
\frac{\lambda_{12}\xi_{2}}{\lambda_{21}\xi_{1}}. \label{gamma-a-ratio}
\end{equation}
The conditions at the external boundaries are $\Phi _{s}^{\prime}(-d_{s})=0$ and
$\Phi_{\alpha}^{\prime}(d_{\mathrm{\pm}})=0$.

The supercurrent flowing through the interface between the s-wave superconductor and
$\alpha$ band is determined by the Green's functions $\Phi_{\{s,\alpha\}}$ at the
interface as
\begin{equation}
j_{\alpha}=\frac{\mathcal{A}_{0}}{\tilde{\gamma}_{B\alpha}}2\pi T\sum
_{\omega>0}\frac{\operatorname{Im}[\Phi_{s}^{\ast}\Phi_{\alpha}]}{\sqrt
{\omega^{2}+|\Phi_{s}|^{2}}\sqrt{\omega^{2}+|\Phi_{\alpha}|^{2}}}
\label{SCorrentIntTxt}
\end{equation}
with $\mathcal{A}_{0}=1/(e\rho_{s}\xi_{s}^{\ast})$. Substitution of the zero-order
approximation for the Green's functions, $\Phi_{\{s,\alpha
\}}\!\approx\!\Delta_{\{s,\alpha\}0}$, gives the well-known Ambegaokar-Baratoff
result \cite{JosCurrTxtbk} for the partial Josephson currents proportional to
$\sin\phi$ with different signs corresponding to the signs of $\Delta_{\alpha0}$.
Here and below we assume for definiteness that $\phi$ is the phase shift between
$\Delta_{s0}$ and $\Delta_{10}$. To find the $\sin(2 \phi)$ term in the Josephson
current one has to go beyond the zero-order approximation and evaluate corrections
to the Green's functions due to the interface.  We discuss these corrections in the
next section.

\section{Proximity corrections in the weak-coupling limit
\label{Sec:Corrections}}

\begin{figure}[h]
\centering
\includegraphics[width=2.5in]{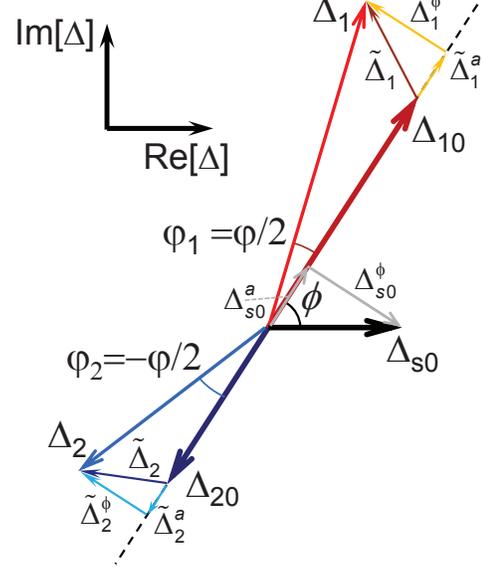}
\caption{Illustration of
the bulk gap parameters, $\Delta_{s0}$ and $\Delta_{\alpha0}$, and
interface-induced corrections to the average $s_{\pm}$ gaps, $\tilde{\Delta
}_{\alpha}$, for a general TRSB state. The gap parameters are presented as
vectors in the complex plane. Decompositions of $\Delta_{s0}$ and
$\tilde{\Delta}_{\alpha}$ into the amplitude and phase components are also
illustrated.}
\label{Fig-DeltasFrustrState}
\end{figure}
In the case of weak coupling between the $s$ and $s_{\pm}$ superconductors,
$\gamma_{B\alpha}\!\gg\!1$, the contact-induced corrections to the gaps and Green's
function can be treated as small perturbations, $\Delta_{\{s,\alpha
\}}(x)=\Delta_{\{s,\alpha\}0}+\tilde{\Delta}_{\{s,\alpha\}}(x)$, $\Phi
_{\{s,\alpha\}}(x)=\Delta_{\{s,\alpha\}0}+\tilde{\Phi}_{\{s,\alpha\}}(x)$. As a
zero-order approximation, we consider a general complex case with a finite phase
difference $\phi$ between the bulk gap parameters $\Delta_{10}$ and $\Delta_{s0}$,
see Fig.\ \ref{Fig-DeltasFrustrState}. For the \emph{aligned} states such
perturbative calculation has been reported in Ref.\
\onlinecite{ProxFingeprintEPL11}. Without loss of generality, we assume
$\Delta_{s0}$ to be real. The small corrections
$\tilde{\Phi}_{\{s,\alpha\}}(\omega,x)$ and $\tilde{\Delta }_{\{s,\alpha\}}(x)$ can
be computed analytically in the linear order with respect to $1/\gamma_{B\alpha}$.
Similar calculation for several types of junctions using somewhat different approach
has been done in Ref.\ \onlinecite{GolubovKuprPZHETF05}. The details of these
derivations are described in Appendix \ref{App-WeakCoupl}. In the TRSB state the
solution for corrections exists only if the partial Josephson energies in the linear
approximation exactly compensate each other, $E_{J1}=E_{J2}$. Here $E_{J\alpha}$ are
related to the gaps and boundary resistance $R_{B\alpha}$ by the standard expression
\begin{equation}
E_{J,\alpha}\!=\!\frac{\hbar}{2e^{2}R_{B}^{\alpha}}2\pi T\sum_{\omega>0}
\frac{\Delta_{s0}|\Delta_{\alpha0}|}{\sqrt{\omega^{2}\!+\!\Delta_{s0}^{2}}
\sqrt{\omega^{2}\!+\!|\Delta_{\alpha0}|^{2}}}. \label{PartJosEner}
\end{equation}
This also means that the total Josephson current flowing through the boundary is
always zero in the ground state. The corrections can be presented in the form of
Fourier expansions. For the $s$-wave superconductor $\tilde{\Phi}
_{s}(\omega,x)\!=\!\sum_{m=0}^{\infty}\tilde{\Phi}_{s,m}(\omega)\cos k_{m} x$,$\
\tilde{\Delta}_{s}(x)\!=\!\sum_{m=0}^{\infty}\tilde{\Delta}_{s,m}\cos k_{m}x$ with
$k_{m}\!=\!m\pi/d_{s}$. The Fourier components of the Green's functions computed in
Appendix \ref{App-weak-s} are given by
\begin{align}
\tilde{\Phi}_{s,m}&\!=\!\frac{\tilde{\Delta}_{s,m}}{1+\xi_{s,\omega}^{2}
k_{m}^{2}}+\frac{(2\!-\!\delta_{m})\xi_{s,\omega}^{2}/(d_{s}\xi_{s}^{\ast}
)}{1+\xi_{s,\omega}^{2}k_{m}^{2}}\nonumber\\
&\times \sum_{\alpha}\frac{\sqrt{\omega
^{2}\!+\!\Delta_{s0}^{2}}}{\sqrt{\omega^{2}\!+\!|\Delta_{\alpha0}|^{2}}}
\frac{\Delta_{\alpha0}\!-\!\Delta_{s0}}{\tilde{\gamma}_{B\alpha}},
\label{PhismTxt}
\end{align}
where
$\xi_{s,\omega}^{2}\!=\!\xi_{s,\Delta}^{2}\Delta_{s0}/\sqrt{\omega^{2}\!+\!\Delta_{s0}^{2}}$,
 $\xi_{s,\Delta}^{2}=D_{s}/(2\Delta_{s0})$, and $\delta_{m}\!=\!1(0)$ for
$m\!=\!0(m\!>\!0)$.  Here the first (bosonic) term is induced by the correction to
the gap parameter and the second (fermionic) term is the direct response to the
boundary perturbation. In the complex state the responses of the gap parameter are
different in the amplitude and phase channels. As $\Delta_{s0}$ is selected real,
these channels correspond to the real and imaginary parts of the gap correction,
$\tilde{\Delta}_{s}=\tilde{\Delta}_{s}^{R}+i\tilde{\Delta}_{s}^{I}$. The Fourier
components of $\tilde{\Delta}_{s}^{R}$ and $\tilde{\Delta}_{s}^{I}$ computed in
Appendix \ref{App-weak-s} can be presented as
\begin{widetext}
\begin{subequations}
\label{CorDeltasTxt}
\begin{align}
\tilde{\Delta}_{s,m}^{R}  &  =\frac{2\pi T}{Z_{s,m}^{a}}\sum_{\alpha,\omega
>0}\frac{\omega^{2}}{\left(  \omega^{2}\!+\!\Delta_{s0}^{2}\right)
\sqrt{\omega^{2}\!+\!|\Delta_{\alpha0}|^{2}}}\frac{\left(  2\!-\!\delta
_{m}\right)  \xi_{s,\Delta}^{2}/\left(  d_{s}\xi_{s}^{\ast}\right)  }
{\sqrt{\omega^{2}\!+\!\Delta_{s0}^{2}}/\Delta_{s0}\!+\!\left(  \pi
m\xi_{s,\Delta}/d_{s}\right)  ^{2}}\frac{\operatorname{Re}[\Delta_{\alpha
0}]\!-\!\Delta_{s0}}{\tilde{\gamma}_{B\alpha}},\label{CorDeltasRTxt}\\
\text{with }  &  Z_{s,m}^{a}=2\pi T\sum_{\omega>0}\frac{1}{(\omega^{2}
+\Delta_{s0}^{2})^{3/2}}\left(  \Delta_{s0}^{2}+\frac{\omega^{2}\left(  \pi
m\xi_{s,\Delta}/d_{s}\right)  ^{2}}{\sqrt{\omega^{2}+\Delta_{s0}^{2}}
/\Delta_{s0}+\left(  \pi m\xi_{s,\Delta}/d_{s}\right)  ^{2}}\right)
\nonumber\\
\tilde{\Delta}_{s,m}^{I}  &  =\frac{2\pi T}{Z_{s,m}^{\phi}}\sum_{\alpha
,\omega>0}\frac{2\xi_{s,\Delta}^{2}/(d_{s}\xi_{s})}{\sqrt{\omega^{2}
+\Delta_{s0}^{2}}/\Delta_{s0}+\left(  \pi m\xi_{s,\Delta}/d_{s}\right)  ^{2}
}\frac{1}{\sqrt{\omega^{2}+|\Delta_{\alpha0}|^{2}}}\frac{\operatorname{Im}
[\Delta_{\alpha0}]}{\tilde{\gamma}_{B\alpha}},\label{CorDeltasITxt}\\
\text{with }  &  Z_{s,m}^{\phi}=2\pi T\sum_{\omega>0}\frac{1}{\sqrt{\omega
^{2}+\Delta_{s0}^{2}}}\frac{\left(  \pi m\xi_{s,\Delta}/d_{s}\right)  ^{2}
}{\sqrt{\omega^{2}+\Delta_{s0}^{2}}/\Delta_{s0}+\left(  \pi m\xi_{s,\Delta
}/d_{s}\right)  ^{2}}.\nonumber
\end{align}
\end{subequations}
\end{widetext}

For the $s_{\pm}$ superconductor the corresponding Fourier series are $\tilde
{\Phi}_{\alpha}\!=\!\sum_{m=0}^{\infty}\tilde{\Phi}_{\alpha,m}\cos q_{m} x$,$\
\tilde{\Delta}_{\alpha}\!=\!\sum_{m=0}^{\infty}\tilde{\Delta}_{\alpha,m}\cos q_{m}x$
with $q_{m}\!=\!m\pi/d_{\mathrm{\pm}}$. Detailed derivations of the Fourier
components are presented in Appendix \ref{App-Weak-spm} and the result for
$\tilde{\Phi}_{\alpha,m}$ can be written as,
\begin{equation}
\tilde{\Phi}_{\alpha,m}=\frac{\tilde{\Delta}_{\alpha,m}}{1+\xi_{\alpha,\omega
}^{2}q_{m}^{2}}+\tilde{\Phi}_{\alpha,b,m}.\label{PhipmTxt}
\end{equation}
where
\[\xi_{\alpha,\omega}^{2}=\xi_{\alpha,\Delta}^{2}|\Delta_{\alpha0}
|/\sqrt{\omega^{2}+|\Delta_{\alpha0}|^{2}},\ \xi_{\alpha,\Delta}^{2}\!=\!
D_{\alpha}\!/(2|\Delta_{\alpha0}|).\]
Here, as in Eq.\ (\ref{PhismTxt}), the first term is induced by the correction to
the gap parameter and the second term is directly induced by the interface. For a
general complex state, the corrections have to be split into the amplitude (along
$\Delta_{\alpha0}$) and phase channels, $\tilde{\Phi}_{\alpha
,b,m}=\tilde{\Phi}_{\alpha,b,m}^{a}+\tilde{\Phi}_{\alpha,b,m}^{\phi}$, with
\begin{align}
\binom{\tilde{\Phi}_{\alpha,b,m}^{a}}{\tilde{\Phi}_{\alpha,b,m}^{\phi}} &
=\frac{\left(  2-\delta_{m}\right)  }{\gamma_{B\alpha}}\frac{\xi
_{\alpha,\omega}^{2}/\left(  d_{\mathrm{\pm}}\xi_{\alpha}\right)  }
{1+\xi_{\alpha,\omega}^{2}q_{m}^{2}}\nonumber\\
&  \times\frac{\sqrt{\omega^{2}+|\Delta_{\alpha0}|^{2}}}{\sqrt{\omega
^{2}+\Delta_{s0}^{2}}}\binom{\Delta_{s0}^{a}\!-\!\Delta_{\alpha0}}{\Delta
_{s0}^{\phi}},\label{PhipmbTxt}
\end{align}
where $\Delta_{s0}^{a}$ ($\Delta_{s0}^{\phi}$) is the projection of $\Delta_{s0}$
along $\Delta_{\alpha0}$ (into perpendicular direction), as illustrated in Fig.
\ref{Fig-DeltasFrustrState}. The gap corrections
$\tilde{\Delta}_{\alpha,m}^{a,\phi}$ are related to $\tilde{\Phi}_{\alpha
,b,m}^{a,\phi}$ as
\begin{subequations}
\label{DPMSolTxt}
\begin{align}
\tilde{\Delta}_{\alpha,m}^{a} &  =2\pi T\sum_{\beta,\omega>0}U_{m,\alpha\beta
}^{a}\frac{\omega^{2}\tilde{\Phi}_{\beta,b,m}^{a}}{\left(  \omega^{2}
+|\Delta_{\beta0}|^{2}\right)  ^{3/2}},\label{DpmaTxt}\\
\tilde{\Delta}_{\alpha,m}^{\phi} &  =2\pi T\sum_{\beta,\omega>0}
U_{m,\alpha\beta}^{\phi}\frac{\tilde{\Phi}_{\beta,b,m}^{\phi}}{\sqrt
{\omega^{2}+|\Delta_{\beta0}|^{2}}},\label{DfpmTxt}
\end{align}
with the matrices $U_{m,\alpha\beta}^{a,\phi}=\left[  w_{\alpha\beta}
-\Sigma_{\alpha,m}^{a,\phi}\delta_{\alpha\beta}\right]  ^{-1}$ where
\end{subequations}
\begin{subequations}
\label{EqSigpm}
\begin{align*}
\Sigma_{\alpha,m}^{a}\! &  =\!2\pi T\!\sum_{\omega>0}\!\left[  \frac
{\omega^{2}}{\left(  \omega^{2}\!+\!|\Delta_{\alpha0}|^{2}\right)
^{3/2}\left(  1\!+\!\xi_{\alpha,\omega}^{2}q_{m}^{2}\right)  }\!-\!\frac
{1}{\omega}\right] \\
& \!+ \ln\frac{T_{c}}{T},\\
\Sigma_{\alpha,m}^{\phi}\!&=\!2\pi T\sum_{\omega>0}\left[  \frac{1}
{\sqrt{\omega^{2}\!+\!|\Delta_{\alpha0}|^{2}}(1\!+\!\xi
_{\alpha,\omega}^{2}q_{m}^{2})}\!-\!\frac{1}{\omega}\right] \\
&\!+ \ln\frac{T_{c}}{T},
\end{align*}
\end{subequations}
$w_{\alpha\beta}=\lambda_{\alpha\beta}^{-1}-\lambda^{-1}\delta_{\alpha\beta}$ and
$\lambda$ is the largest eigenvalue of the matrix $\lambda_{\alpha\beta}$. In the
two-band case the explicit formulas for $w_{\alpha\beta}$ and
$U_{m,\alpha\beta}^{a,\phi}$ are given in the Appendix \ref{App-Weak-spm}, Eqs.
(\ref{wab}) and (\ref{Uaphiab}).

In summary, Eqs.\ (\ref{PhismTxt}) and (\ref{CorDeltasTxt}) give corrections to the
gap parameters and Green's function for the $s$-wave superconductor while Eqs.\
(\ref{PhipmTxt}), (\ref{PhipmbTxt}), and (\ref{DPMSolTxt}) give corresponding
results for the $s_{\pm}$ superconductor. These corrections will allow us to derive
in the next section a general result for the second harmonic of the Josephson
current that determines the width of the TRSB region.

\section{Second harmonic of the Josephson current \label{Sec:SecondHarm}}

We already mentioned that the linear order with respect to the coupling strength
$\propto 1/\tilde{\gamma }_{B\alpha}$ is not sufficient to determine the range of
parameters where the TRSB state is realized. As discussed in Sec.\
\ref{Sec:TransSecondHarm}, in the weak-coupling regime this range is determined by
the term $\propto \sin(2\phi)$ in the Josephson current that appears only in the
quadratic order. In this section we derive microscopic expression for this term
using corrections to the Green's functions presented in the previous section.

For arbitrary coupling the current density flowing through the interface between the
s-wave superconductor and $\alpha$-band is given by Eq.\ (\ref{SCorrentIntTxt}) and
in ground state
\[
j=\sum_{\alpha}j_{\alpha}=\mathcal{A}_{0}\mathcal{I}=0.
\]
Here the parameter $\mathcal{I}$ has dimensionality of energy. To find the
second-order term in $j$, we have to expand the right hand side of Eq.\
(\ref{SCorrentIntTxt}) with respect to small corrections to $\Phi_{s}$ and
$\Phi_{\alpha}$. This gives $\mathcal{I\approx I}^{(1)}+\mathcal{I}^{(2)}$ where the
term
\begin{equation}
\mathcal{I}^{(1)}\!  =2\pi T\!\sum_{\alpha,\omega>0}\frac{\operatorname{Im}
[\Delta_{s0}^{\ast}\Delta_{\alpha0}]}{\tilde{\gamma}_{B\alpha}\sqrt{\omega
^{2}\!+\!|\Delta_{\alpha0}|^{2}}\sqrt{\omega^{2}+|\Delta_{s0}|^{2}}
}\!\label{I1}
\end{equation}
corresponds to the standard main-order Josephson current and
\begin{widetext}
\begin{equation}
\mathcal{I}^{(2)}\!   =\!2\pi T\!\sum_{\alpha,\omega>0}\!\frac
{\operatorname{Im}[\tilde{\Phi}_{s}^{\ast}\Delta_{\alpha0}\!+\!\Delta
_{s0}^{\ast}\tilde{\Phi}_{\alpha}]\!-\!\operatorname{Im}[\Delta_{s0}^{\ast
}\Delta_{\alpha0}]\left(  \frac{\operatorname{Re}[\Delta_{s0}^{\ast}
\tilde{\Phi}_{s}]}{\omega^{2}\!+\!|\Delta_{s0}|^{2}}\!+\!\frac
{\operatorname{Re}[\Delta_{\alpha0}^{\ast}\tilde{\Phi}_{\alpha}]}{\omega
^{2}\!+\!|\Delta_{\alpha0}|^{2}}\right)  }{\tilde{\gamma}_{B\alpha}
\sqrt{\omega^{2}\!+\!|\Delta_{\alpha0}|^{2}}\sqrt{\omega^{2}\!+\!|\Delta
_{s0}|^{2}}} \label{I2}
\end{equation}
\end{widetext}
is the second-order term which is determined by the linear corrections to the
Green's functions due to the interface perturbations, $\tilde{\Phi}_{\{s,\alpha\}}$,
considered in the previous section. Using these results, we can present the
corrections at $x=0$ in the form
\begin{subequations}
\begin{align}
\tilde{\Phi}_{s}^{R}(0)  &  =\!\sum_{\alpha}\mathcal{F}_{s,\alpha}^{a}
\frac{\Delta_{s0}+(-1)^{\alpha}|\Delta_{\alpha0}|\cos\phi}{\tilde{\gamma
}_{B\alpha}},\ \label{Phi-s-Fa}\\
\tilde{\Phi}_{s}^{I}(0)  &  =\!\sum_{\alpha}\mathcal{F}_{s,\alpha}^{\phi}
\frac{(-1)^{\alpha}|\Delta_{\alpha0}|\sin\phi}{\tilde{\gamma}_{B\alpha}
}\label{Phi-s-Fphi}\\
\tilde{\Phi}_{\alpha}^{a}(0)  &  =\!\sum_{\beta}\mathcal{F}_{\alpha\beta}
^{a}\frac{\Delta_{s0}^{a}\!-\!\Delta_{\beta0}}{\gamma_{B\beta}},\ \tilde{\Phi
}_{\alpha}^{\phi}(0)\!=\!\sum_{\beta}\mathcal{F}_{\alpha\beta}^{\phi}
\frac{\Delta_{s0}^{\phi}}{\gamma_{B\beta}}, \label{Phi-ab-F}
\end{align}
where the response functions of the $s$-wave superconductor in the amplitude and
phase channels, $\mathcal{F}_{s,\alpha}^{a,\phi}(\omega)$, can be explicitly written
as
\end{subequations}
\begin{subequations}
\label{Fsa}
\begin{align}
\mathcal{F}_{s,\alpha}^{a}  &  =\!-\frac{\xi_{s,\omega}/\xi_{s}^{\ast}}
{\tanh\left(  d_{s}/\xi_{s,\omega}\right)  }\frac{\sqrt{\omega^{2}+\Delta
_{s0}^{2}}}{\sqrt{\omega^{2}+|\Delta_{\alpha0}|^{2}}}-\sum_{m=0}^{\infty}
\frac{2-\delta_{m}}{1+\xi_{s,\omega}^{2}k_{m}^{2}}\nonumber\\
&  \times\frac{2\pi T}{Z_{s,m}^{a}}\sum_{\omega_{1}>0}\frac{\omega_{1}^{2}
}{\left(  \omega_{1}^{2}+\Delta_{s0}^{2}\right)  \sqrt{\omega_{1}^{2}
+|\Delta_{\alpha0}|^{2}}}\frac{\xi_{s,\omega_{1}}^{2}/\left(  d_{s}\xi
_{s}^{\ast}\right)  }{1+\xi_{s,\omega_{1}}^{2}k_{m}^{2}},\label{Fsa-a}\\
\mathcal{F}_{s,\alpha}^{\phi}  &  =-\frac{\xi_{s,\omega}/\xi_{s}^{\ast}}
{\tanh\left(  d_{s}/\xi_{s,\omega}\right)  }\frac{\sqrt{\omega^{2}+\Delta
_{s0}^{2}}}{\sqrt{\omega^{2}+|\Delta_{\alpha0}|^{2}}}-\sum_{m=1}^{\infty}
\frac{2}{1+\xi_{s,\omega}^{2}k_{m}^{2}}\nonumber\\
&  \times\frac{2\pi T}{Z_{s,m}^{\phi}}\sum_{\omega_{1}>0}\frac{1}{\sqrt
{\omega_{1}^{2}+|\Delta_{\alpha0}|^{2}}}\frac{\xi_{s,\omega_{1}}^{2}/\left(
d_{s}\xi_{s}^{\ast}\right)  }{1+\xi_{s,\omega_{1}}^{2}k_{m}^{2}},
\label{Fsa-phi}
\end{align}
with $Z_{s,m}^{a,\phi}$ defined in Eq.\ (\ref{CorDeltasTxt}). The response functions
of the $s_{\pm}$ superconductor, $\mathcal{F}_{\alpha\beta}^{a,\phi}(\omega)$, are
given by
\end{subequations}
\begin{subequations}
\label{Fab}
\begin{align}
&  \mathcal{F}_{\alpha\beta}^{a}\!=\!\frac{\xi_{\alpha,\omega}/\xi_{\alpha}}
{\tanh\left(  d_{\mathrm{\pm}}/\xi_{\alpha,\omega}\right)  }\frac{\sqrt
{\omega^{2}\!+\!|\Delta_{\alpha0}|^{2}}}{\sqrt{\omega^{2}\!+\!\Delta_{s0}^{2}}}
\delta_{\alpha\beta}\!+\!\sum_{m=0}^{\infty}\frac{2-\delta_{m}}{1\!+\!\xi
_{\alpha,\omega}^{2}q_{m}^{2}}\nonumber\\
&  \times\! 2\pi T\sum_{\omega_{1}>0}\!U_{m,\alpha\beta}^{a}\frac{\omega_{1}^{2}
}{\left(  \omega_{1}^{2}\!+\!|\Delta_{\beta0}|^{2}\right)  \sqrt{\omega_{1}
^{2}\!+\!\Delta_{s0}^{2}}}\frac{\xi_{\beta,\omega_{1}}^{2}/\left( d_{\mathrm{\pm
}}\xi_{\beta}\right)  }{1\!+\!\xi_{\beta,\omega_{1}}^{2}q_{m}^{2}},\label{Fab-a}\\
&  \mathcal{F}_{\alpha\beta}^{\phi}=\mathcal{F}_{\alpha,0}^{\phi}
\delta_{\alpha\beta}+\frac{\xi_{\alpha,\omega}/\xi_{\alpha}}{\tanh\left(
d_{\mathrm{\pm}}/\xi_{\alpha,\omega}\right)  }\frac{\sqrt{\omega^{2}
+|\Delta_{\alpha0}|^{2}}}{\sqrt{\omega^{2}+\Delta_{s0}^{2}}}\delta
_{\alpha\beta}\nonumber\\
&  +\!\sum_{m=1}^{\infty}\frac{2}{1\!+\!\xi_{\alpha,\omega}^{2}q_{m}^{2}}2\pi
T\sum_{\omega_{1}>0}\!U_{m,\alpha\beta}^{\phi}\frac{1}{\sqrt{\omega_{1}
^{2}\!+\!\Delta_{s0}^{2}}}\frac{\xi_{\beta,\omega_{1}}^{2}/\left(  d_{\mathrm{\pm
}}\xi_{\beta}\right)  }{1\!+\!\xi_{\beta,\omega_{1}}^{2}q_{m}^{2}},
\label{Fab-phi}
\end{align}
where the matrices $U_{m,\alpha\beta}^{a,\phi}$ are defined in Eq. (\ref{Uaphiab}),
and the term
\end{subequations}
\begin{equation}
\mathcal{F}_{\alpha,0}^{\phi}=\frac{2\pi T}{2d_{\mathrm{\pm}}w_{12}
|\Delta_{20}|}\sum_{\omega>0}\frac{\xi_{1,\omega}^{2}/\xi_{1}}{\sqrt
{\omega^{2}+\Delta_{s0}^{2}}}|\Delta_{\alpha0}| \label{Fab-phi0}
\end{equation}
describes the contribution from the uniform phase corrections $\tilde{\Delta
}_{\alpha,0}^{\phi}$ given by Eq.\ (\ref{dPhi0}). The different quantities entering
Eq. (\ref{I2}) can now be expressed as
\begin{align*}
\operatorname{Im}[\Delta_{s0}^{\ast}\Delta_{\alpha0}]  &=-\Delta_{s0}
|\Delta_{\alpha0}|(-1)^{\alpha}\sin\phi,\\
\operatorname{Im}[\tilde{\Phi}_{s}^{\ast}\Delta_{\alpha0}]  &=|\Delta
_{\alpha0}|\sum_{\beta}\frac{1}{\tilde{\gamma}_{B\beta}}\left[  -\mathcal{F}
_{s,\beta}^{a}\Delta_{s0}(-1)^{\alpha}\sin\phi\right. \\
-&   \left.  \left(  \mathcal{F}_{s,\beta}^{a}-\mathcal{F}_{s,\beta}^{\phi
}\right)  |\Delta_{\beta0}|(-1)^{\alpha+\beta}\sin\phi\cos\phi\right]  ,\\
\operatorname{Im}[\Delta_{s0}^{\ast}\tilde{\Phi}_{\alpha}]  &  =\Delta
_{s0}\sum_{\beta}\frac{1}{\gamma_{B\beta}}\left[  (-1)^{\beta}\mathcal{F}
_{\alpha\beta}^{a}|\Delta_{\beta0}|\sin\phi\right. \\
+&   \left.  \left(  \mathcal{F}_{\alpha\beta}^{a}-\mathcal{F}_{\alpha\beta
}^{\phi}\right)  \Delta_{s0}\sin\phi\cos\phi\right]  ,\\
\operatorname{Re}[\Delta_{s0}^{\ast}\tilde{\Phi}_{s}^{a}]  &  =\Delta_{s}
\sum_{\beta}\mathcal{F}_{s,\beta}^{a}\frac{\Delta_{s0}+(-1)^{\beta}
|\Delta_{\beta0}|\cos\phi}{\tilde{\gamma}_{B\beta}},\\
\operatorname{Re}[\Delta_{\alpha0}^{\ast}\tilde{\Phi}_{\alpha}]\!
=&\!-\!(-1)^{\alpha}|\Delta_{\alpha0}|\!\sum_{\beta}\mathcal{F}_{\alpha\beta
}^{a}\frac{\Delta_{s0}\cos\phi\!+\!(-1)^{\beta}|\Delta_{\beta0}|}
{\gamma_{B\beta}}.
\end{align*}
We can see that $\mathcal{I}^{(2)}$ contains terms proportional to $\sin\phi$ and
$\sin\phi\cos\phi\!=\!\frac{1}{2}\sin(2\phi)$, $\mathcal{I}^{(2)}
\mathcal{=J}^{(1)}\sin\phi+\mathcal{J}^{(2)}\cos\phi\sin\phi$. The terms
$\propto\sin\phi$ just slightly shift location of the transition line. The
transition order and possible width of the TRSB\ region are determined by the terms
$\propto\sin\phi\cos\phi$. Collecting such terms in $\mathcal{I}$, we obtain
\begin{align}
\mathcal{J}^{(2)}=  &  2\pi T\!\sum_{\alpha,\omega>0}\!\frac{|\Delta_{\alpha
0}|\Delta_{s0}\mathcal{R}_{\alpha}(\omega)}{\tilde{\gamma}_{B\alpha}
\sqrt{\omega^{2}\!+\!|\Delta_{\alpha0}|^{2}}\sqrt{\omega^{2}\!+\!\Delta
_{s0}^{2}}}\label{J2}\\
\mathcal{R}_{\alpha}(\omega)  &  =\sum_{\beta}\left[  -\frac{|\Delta_{\beta
0}|(-1)^{\alpha+\beta}}{\tilde{\gamma}_{B\beta}\Delta_{s0}}\left(
\frac{\omega^{2}\mathcal{F}_{s,\beta}^{a}}{\omega^{2}+\Delta_{s0}^{2}
}-\mathcal{F}_{s,\beta}^{\phi}\right)  \right. \nonumber\\
&  \left.  +\frac{\Delta_{s0}}{\gamma_{B\beta}|\Delta_{\alpha0}|}\left(
\frac{\omega^{2}\mathcal{F}_{\alpha\beta}^{a}}{\omega^{2}+|\Delta_{\alpha
0}|^{2}}-\mathcal{F}_{\alpha\beta}^{\phi}\right)  \right] \nonumber
\end{align}
The term corresponding to the uniform phase tilt $\mathcal{F}_{\alpha,0} ^{\phi}$,
Eq. (\ref{Fab-phi0}) has special meaning and it is useful to evaluate it explicitly,
\begin{align}
\mathcal{J}_{\phi,0}^{(2)}  &  =\!-\!\left(  \frac{2\pi T}{\tilde{\gamma}
_{B1}}\!\sum_{\omega>0}\!\frac{\Delta_{s0}|\Delta_{10}|}{\sqrt{\omega
^{2}\!+\!|\Delta_{10}|^{2}}\sqrt{\omega^{2}\!+\!|\Delta_{s0}|^{2}}}\right)
^{2}\nonumber\\
&  \times\frac{\gamma_{1}\xi_{1,\Delta}^{2}|\Delta_{10}|/\xi_{1}
}{d_{\mathrm{\pm}}w_{12}|\Delta_{10}||\Delta_{20}|}\nonumber\\
&  =-\frac{2e^{2}}{\hbar}E_{J,1}^{2}\frac{\rho_{s}\xi_{s}}{d_{\mathrm{\pm}}
\nu_{1}w_{12}|\Delta_{10}||\Delta_{20}|}, \label{J2phi0}
\end{align}
where we used the relations $\gamma_{1}\xi_{1,\Delta}^{2}|\Delta_{10}|/\xi
_{1}=\hbar\rho_{1}D_{1}/(2\rho_{s}\xi_{s})=\hbar/(2e^{2}\nu_{1}\rho_{s}\xi _{s})$.
This gives the following contribution to the second harmonic of the Josephson
current
\begin{equation}
j_{\phi,0}^{(2)}=-j_{J,1}\frac{E_{J,1}}{2d_{\mathrm{\pm}}\nu_{1}w_{12}
|\Delta_{10}||\Delta_{20}|}. \label{j2phi0}
\end{equation}
The quantity $\mathcal{E}_{12}=d_{\mathrm{\pm}}\nu_{1}w_{12}
|\Delta_{10}||\Delta_{20}|$ in the denominator represents the interband coupling
energy. Therefore this result exactly corresponds to the result of the frustrated
Josephson junction model. This term, however, dominates only in the case of small
interband coupling energy, when the parameter $w_{12}$ is very small.

To gain a further insight on the structure of the second-harmonic amplitude
$\mathcal{J}^{(2)}$, we present it explicitly as a sum of terms corresponding to
contributions from the corrections to the $s$-wave and $s_{\pm}$ Green's functions
coming directly from the interface ($\mathcal{J}_{b,\ast}^{(2)}$) and via gap
parameters ($\mathcal{J}_{\Delta,\ast}^{(2)}$).
\begin{equation}
\mathcal{J}^{(2)}=\mathcal{J}_{b,s}^{(2)}+\mathcal{J}_{\Delta,s}
^{(2)}+\mathcal{J}_{b,\mathrm{pm}}^{(2)}+\mathcal{J}_{\Delta,\mathrm{pm}}^{(2)}+\mathcal{J}
_{\phi,0}^{(2)} \label{SumTerms}
\end{equation}
We already considered above the last term, $\mathcal{J}_{\phi,0}^{(2)}$, that is
part of $\mathcal{J} _{\Delta,\mathrm{pm}}^{(2)}$ coming the uniform phase response
of $s_{\pm}$ superconductor. This term requires separate treatment leading to Eq.\
(\ref{J2phi0}) which corresponds to the result of the frustrated Josephson junction
model. As for the other terms, the $s$-wave components are given by the following
explicit formulas,
\begin{widetext}
\begin{subequations}
\label{J2s}
\begin{align}
\mathcal{J}_{b,s}^{(2)}  &  =-2\pi T\sum_{\omega>0}\left(  \sum_{\alpha}
\frac{\Delta_{\alpha0}}{\tilde{\gamma}_{B\alpha}\sqrt{\omega^{2}+|\Delta
_{\alpha0}|^{2}}}\right)  ^{2}\frac{\Delta_{s0}^{2}}{\omega^{2}+\Delta_{s0}^{2}
}\frac{\xi_{s,\omega}/\xi_{s}}{\tanh\left(  d_{s}/\xi_{s,\omega}\right)
},\label{J2bs}\\
\mathcal{J}_{\Delta,s}^{(2)}  &  =-\frac{\xi_{s,\Delta}^{2}}{\Delta_{s0}
d_{s}\xi_{s}}\left[  -\frac{Y_{a,0}^{2}}{Z_{s,0}^{a}}+2\sum_{m=1}^{\infty
}\left(  \frac{Y_{\phi,m}^{2}}{Z_{s,m}^{\phi}}-\frac{Y_{a,m}^{2}}{Z_{s,m}^{a}
}\right)  \right]  ,\label{J2Ds}\\
Y_{\phi,m}  &  =2\pi T\sum_{m,\alpha,\omega>0}\frac{\Delta_{\alpha0}}
{\tilde{\gamma}_{B\alpha}\sqrt{\omega^{2}+|\Delta_{\alpha0}|^{2}}\left(
\sqrt{\omega^{2}+\Delta_{s0}^{2}}/\Delta_{s0}+\xi_{\Delta,s}^{2}k_{m}
^{2}\right)  },\nonumber\\
Y_{a,m}  &  =2\pi T\sum_{m,\alpha,\omega>0}\frac{\Delta_{\alpha0}\omega^{2}
}{\tilde{\gamma}_{B\alpha}\sqrt{\omega^{2}+|\Delta_{\alpha0}|^{2}}\left(
\omega^{2}+\Delta_{s0}^{2}\right)  \left(  \sqrt{\omega^{2}+\Delta_{s0}^{2}
}/\Delta_{s0}+\xi_{\Delta,s}^{2}k_{m}^{2}\right)  },\nonumber
\end{align}
while the $s_{\pm}$ components can be written as
\end{subequations}
\begin{subequations}
\label{J2pm}
\begin{align}
\mathcal{J}_{b,\mathrm{pm}}^{(2)}  &  =-2\pi T\sum_{\alpha,\omega>0}\frac{1}
{\tilde{\gamma}_{B\alpha}\gamma_{B\alpha}}\frac{\Delta_{s0}^{2}
|\Delta_{\alpha0}|^{2}}{\left(  \omega^{2}+\Delta_{s0}^{2}\right)  \left(  \omega^{2}
+|\Delta_{\alpha0}|^{2}\right)  }\frac{\xi_{\alpha,\omega}/\xi_{\alpha}}
{\tanh\left(  d_{\mathrm{\pm}}/\xi_{\alpha,\omega}\right)  },\label{J2bpm}\\
\mathcal{J}_{\Delta,\mathrm{pm}}^{(2)}  &  =-\frac{\pi T_{c}}{d_{\mathrm{\pm}}}
\Delta_{s0}^{2}\sum_{\alpha,\beta}\left[  -X_{0,\alpha}^{a}U_{0,\alpha\beta
}^{a}\gamma_{\beta}\xi_{\beta}X_{0,\beta}^{a}+2\sum_{m=1}^{\infty}\left(
X_{m,\alpha}^{\phi}U_{m,\alpha\beta}^{\phi}\gamma_{\beta}\xi_{\beta}
X_{m,\beta}^{\phi}-X_{m,\alpha}^{a}U_{m,\alpha\beta}^{a}\gamma_{\beta}
\xi_{\beta}X_{m,\alpha}^{a}\right)  \right]  ,\label{J2Dpm}\\
X_{m,\alpha}^{\phi}  &  =\frac{2\pi T}{\tilde{\gamma}_{B\alpha}}\sum
_{\omega>0}\frac{1}{\sqrt{\omega^{2}+\Delta_{s0}^{2}}\left(  \sqrt{\omega
^{2}+|\Delta_{\alpha0}|^{2}}+|\Delta_{\alpha0}|\xi_{\Delta,\alpha}^{2}q_{m}
^{2}\right)  },\nonumber\\
X_{m,\alpha}^{a}  &  =\frac{2\pi T}{\tilde{\gamma}_{B\alpha}}\sum_{\omega
>0}\frac{\omega^{2}}{\sqrt{\omega^{2}+\Delta_{s0}^{2}}\left(  \omega
^{2}+|\Delta_{\alpha0}|^{2}\right)  \left(  \sqrt{\omega^{2}+|\Delta_{\alpha0}|^{2}}
+|\Delta_{\alpha0}|\xi_{\Delta,\alpha}^{2}q_{m}^{2}\right)  }.\nonumber
\end{align}
\end{subequations}
\end{widetext}
We note also a useful relation for the combination $\gamma_{\beta}\xi_{\beta}$
entering Eq. (\ref{J2Dpm}), $\gamma_{\beta}\xi_{\beta}=\xi_{s}^{\ast}\nu
_{s}/\nu_{\beta}$. In summary, Eqs.\ (\ref{J2phi0}), (\ref{J2s}), and (\ref{J2pm})
give general expressions for the components contributing to the second harmonic of
Josephson current in Eq.\ (\ref{SumTerms}). Even though these formulas are rather
cumbersome, they are suitable for numerical evaluation of the second-harmonic
amplitude for arbitrary parameters of superconductors and interface. In the next
section we analyze these terms for practically important particular case in which
much simpler analytical results can be derived.

\subsection{Analysis of terms for low temperatures in the case $d_{s}<\xi_{s}$
and $\Delta_{s}\ll|\Delta_{\alpha}|$}

Unfortunately, general formulas derived in the previous section are rather
cumbersome. To understand better the relation between different terms and their
absolute values, in this section we evaluate them at low temperature and for the
most interesting case of weaker $s$-wave superconductor, small $d_{s}$ and large
$d_{\mathrm{\pm}}$. In these limits it is possible to derive simple analytical
results for the most important terms.

\subsubsection{Terms $\mathcal{J}_{b,s}^{(2)}$ and $\mathcal{J}_{\Delta
,s}^{(2)}$}

For very thin $s$-wave superconductor the dominating in $1/d_{s}$ order term is
coming from $\mathcal{J}_{b,s}^{(2)}$ and the $m\!=\!0$ term in the amplitude part
of $\mathcal{J}_{\Delta,s}^{(2)}$ which we will notate as $\mathcal{J}
_{\Delta,s,0}^{(2)}$,
\begin{align*}
\mathcal{J}_{b,s}^{(2)}\!  &  =\!-\frac{1}{d_{s}}\int\limits_{0}^{\infty
}\!d\omega\!\left(  \sum_{\alpha}\!\frac{\Delta_{\alpha0}}{\tilde{\gamma
}_{B\alpha}\sqrt{\omega^{2}\!+\!|\Delta_{\alpha0}|^{2}}}\right)^{2}
\!\frac{\Delta_{s0}^{2}}{(\omega^{2}\!+\!\Delta_{s0}^{2})^{3/2}},\\
\mathcal{J}_{\Delta,s,0}^{(2)}\!  &  \approx\!\frac{1}{d_{s}}\left(
\!\sum_{\alpha}\int\limits_{0}^{\infty}d\omega\frac{\Delta_{\alpha0}
\Delta_{s0}^{2}}{\tilde{\gamma}_{B\alpha}\sqrt{\omega^{2}\!+\!|\Delta
_{\alpha0}|^{2}}\left(  \omega^{2}\!+\!\Delta_{s0}^{2}\right)  ^{3/2}
}\!\right)  ^{2},
\end{align*}
where in the last formula we used a compensation condition for the Josephson energy
near the transition point. In the limit $\Delta_{s0}\ll|\Delta _{\alpha0}|$
integrals converge at $\omega\sim\Delta_{s0}$ and one may think than it is possible
to replace $\frac{|\Delta_{\alpha0}|}{\sqrt{\omega
^{2}+|\Delta_{\alpha0}|^{2}}}\rightarrow1$ under the frequency integrals. However,
as $\int_{0}^{\infty}d\omega\frac{\Delta_{s0}^{2}}{(\omega^{2}
+\Delta_{s0}^{2})^{3/2}}=1$, within this approximation the two terms in the sum
\[
\mathcal{J}_{0,s}^{(2)}=\mathcal{J}_{b,s}^{(2)}+\mathcal{J}_{\Delta
,s,0}^{(2)}
\]
exactly compensate each other,
\[
\binom{\mathcal{J}_{b,s}^{(2)}}{\mathcal{J}_{\Delta,s,0}^{(2)}}\approx\mp
\frac{1}{d_{s}}\left(  \sum_{\alpha}\frac{(-1)^{\alpha}}{\tilde{\gamma
}_{B\alpha}}\right)  ^{2},
\]
and therefore they must be evaluated in higher order with respect to
$\Delta_{s0}/|\Delta_{\alpha0}|$. To proceed, we introduce the definitions
\[
L(\omega)  =\sum_{\alpha}\frac{(-1)^{\alpha}|\Delta_{\alpha0}|}
{\tilde{\gamma}_{B\alpha}\sqrt{\omega^{2}+|\Delta_{\alpha0}|^{2}}}=L_{0}
+L_{1}(\omega),
\]
with $L_{0} =\sum_{\alpha}(-1)^{\alpha}/\tilde{\gamma }_{B\alpha}$ and
\[
L_{1}(\omega)  =\sum_{\alpha}\frac{(-1)^{\alpha}}{\tilde{\gamma}_{B\alpha}
}\left(  \frac{|\Delta_{\alpha0}|}{\sqrt{\omega^{2}+|\Delta_{\alpha0}|^{2}}
}-1\right),
\]
which allows us to represent
\begin{align*}
\mathcal{J}_{b,s}^{(2)}  &  =-\frac{1}{d_{s}}\int_{0}^{\infty}d\omega
L^{2}(\omega)\frac{\Delta_{s0}^{2}}{(\omega^{2}+\Delta_{s0}^{2})^{3/2}},\\
\mathcal{J}_{\Delta,s,0}^{(2)}  &  =\frac{1}{d_{s}}\left(  \int_{0}^{\infty
}d\omega L(\omega)\frac{\Delta_{s0}^{2}}{\left(  \omega^{2}+\Delta_{s0}
^{2}\right)  ^{3/2}}\right)  ^{2},
\end{align*}
and rewrite $\mathcal{J}_{0,s}^{(2)}$ as
\begin{align*}
\mathcal{J}_{0,s}^{(2)}  &  =\frac{1}{d_{s}}\left(  \int_{0}^{\infty}d\omega
L_{1}(\omega)\frac{\Delta_{s}^{2}}{\left(  \omega^{2}+\Delta_{s}^{2}\right)
^{3/2}}\right)  ^{2}\\
&  -\frac{1}{d_{s}}\int_{0}^{\infty}d\omega L_{1}^{2}(\omega)\frac{\Delta
_{s}^{2}}{(\omega^{2}+\Delta_{s}^{2})^{3/2}}.
\end{align*}
The dominating contribution is coming from the second term and evaluating integral,
we finally obtain
\begin{align}
&  \mathcal{J}_{0,s}^{(2)} \approx-\frac{\Delta_{s}^{2}}{d_{s}}\left[
\frac{2\ln2-1}{2}\left(  \sum_{\alpha}\frac{\left(  -1\right)  ^{\alpha}
}{\tilde{\gamma}_{B\alpha}|\Delta_{\alpha}|}\right)  ^{2}\right. \nonumber\\
&  -\left.  \frac{\frac{1}{|\Delta_{1}|^{2}}\ln\left(  1+\frac{|\Delta_{1}
|}{|\Delta_{2}|}\right)  +\frac{1}{|\Delta_{2}|^{2}}\ln\left(  1+\frac
{|\Delta_{2}|}{|\Delta_{1}|}\right)  -\frac{2\ln2}{|\Delta_{1}||\Delta_{2}|}
}{\tilde{\gamma}_{B1}\tilde{\gamma}_{B2}}\right]  \label{J20sids}
\end{align}
It is interesting to note that this $1/d_{s}$ term only exists in the asymmetric
case, it vanishes for identical $s_{\pm}$ bands.

\subsubsection{Term $\mathcal{J}_{b,\mathrm{pm}}^{(2)}$}

The term $\mathcal{J}_{b,\mathrm{pm}}^{(2)}$ in Eq.\ (\ref{J2bpm}) for $T=0$ and
$d_{\mathrm{\pm} }\gg\xi_{\alpha}$ becomes
\begin{align}
\mathcal{J}_{b,\mathrm{pm}}^{(2)}  &  \!=\!-\sqrt{\pi T_{c}}\sum_{\alpha}
\frac{\Delta_{s0}^{2} |\Delta_{\alpha0}|^{2}}{\tilde{\gamma}_{B\alpha}
\gamma_{B\alpha}}\nonumber\\
&  \times\int_{0}^{\infty} \!\frac{d\omega}{\left(  \omega^{2}+\Delta_{s0}
^{2}\right)  \left(  \omega^{2}\!+\!|\Delta_{\alpha0}|^{2}\right)  ^{5/4}}.
\label{J2bpmT0}
\end{align}
For $\Delta_{s0}\ll|\Delta_{\alpha0}|$ we can evaluate the frequency integral
leading to the quite simple result
\begin{equation}
\mathcal{J}_{b,\mathrm{pm}}^{(2)}\approx-\frac{\pi}{2}\sum_{\alpha}
\frac{\Delta_{s0} }{\tilde{\gamma}_{B\alpha}\gamma_{B\alpha}}\sqrt{\frac{\pi
T_{c}} {|\Delta_{\alpha0}|}}. \label{J2bpmResult}
\end{equation}
In most cases \emph{this is actually a dominating term} which may be used for an
approximate evaluation of the total second-harmonic amplitude. It has only linear
order with respect to $\Delta_{s0}$, while other terms are proportional to
$\Delta_{s0}^{2}$. Also, for a typical contact $\gamma_{\alpha}=\tilde
{\gamma}_{B\alpha}/\gamma_{B\alpha}\gg1$ due to semimetalic nature of iron-based
superconductors, which enhances $\mathcal{J}_{\ast,\mathrm{pm} }^{(2)}$ terms in
comparison with $\mathcal{J}_{\ast,s}^{(2)}$ terms. The negative sign of
$\mathcal{J}_{b,\mathrm{pm}}^{(2)}$ implies the continuous-transition scenario and
the existence of the TRSB state. In particular, comparing $\mathcal{J}
_{b,\mathrm{pm}}^{(2)}$ with term $\mathcal{J}_{0,s}^{(2)}$, Eq. (\ref{J20sids}), we
obtain, up to a dimensionless function of the ratios $\tilde{\gamma}_{B1}
/\tilde{\gamma}_{B2}$ and $|\Delta_{10}|/|\Delta_{20}|$,
\[
\frac{\mathcal{J}_{b,\mathrm{pm}}^{(2)}}{\mathcal{J}_{0,s}^{(2)}}\approx
\gamma_{1}\frac{d_{s}}{\xi_{s}^{\ast}}\frac{|\Delta_{10}|}{\Delta_{s0}},
\]
which means that the $s$-wave term $\mathcal{J}_{0,s}^{(2)}$ exceeds
$\mathcal{J}_{b,\mathrm{pm}}^{(2)}$ only for an extremely thin $s$-wave layer $d_{s}
<\xi_{s}^{\ast}\Delta_{s0}/\left(  \gamma_{1}|\Delta_{10}|\right)  $. Another factor
further enhancing this ratio is that different $s_{\pm}$ bands contribute to
$\mathcal{J}_{b,\mathrm{pm}}^{(2)}$ with the same sign while their contributions to
$\mathcal{J}_{0,s}^{(2)}$ partially compensate one another.

\subsubsection{Term $\mathcal{J}_{\Delta,\mathrm{pm}}^{(2)}$}

Finally, we obtain the limiting form for the most complicated term
$\mathcal{J}_{\Delta,\mathrm{pm}}^{(2)}$ in Eq.\ (\ref{J2Dpm}). The quantities
$\Sigma_{m,\alpha}^{a,\phi}$ that determine the matrices $U_{m,\alpha\beta
}^{a,\phi}$ in Eq. (\ref{Uaphiab}) can be evaluated at $T=0$ exactly:
\begin{subequations}
\begin{align}
\Sigma_{m,\alpha}^{a,\phi}  &  =g_{a,\phi}\left(  a_{m}\right)  -\ln\left(
\frac{|\Delta_{\alpha0}|}{4\pi T_{c}}\right)  +\psi(1/2),\label{ZaphimT0}\\
g_{\phi}(a)  &  =-\frac{2a}{\sqrt{1-a^{2}}}\arctan\sqrt{\frac{1-a}{1+a}},\\
g_{a}(a)  &  =-\frac{2}{a}\left(  \frac{\pi}{4}-\sqrt{1-a^{2}}\arctan
\sqrt{\frac{1-a}{1+a}}\right)  ,
\end{align}
where $a_{m}=\left(  \pi m\xi_{\alpha,\Delta}/d_{\mathrm{\pm}}\right)  ^{2}$ and
$\psi(x)$ is the digamma function.  The quantities $X_{m,\alpha}^{a,\phi}$ can be
evaluated approximately in the limit $\Delta_{s0}\ll|\Delta_{\alpha0}|$,
\end{subequations}
\begin{subequations}
\begin{align}
X_{m,\alpha}^{\phi}  & \!\approx\!\frac{1}{\tilde{\gamma}_{B\alpha}
|\Delta_{\alpha0}|}\frac{1}{1\!+\!a_{m}}\left(  \ln\frac{4|\Delta_{\alpha0}|}
{\Delta_{s0}}\!+\!\frac{a_{m}}{1\!-\!a_{m}}\ln\frac{2}{1\!+\!a_{m}}\right)
,\label{XmphiT0}\\
X_{m,\alpha}^{a}  &  \approx\frac{1}{\tilde{\gamma}_{B\alpha}|
\Delta_{\alpha0}|}\frac{\ln(1+a_{m})}{a_{m}}. \label{XmaT0}
\end{align}
Assuming that the values for $\xi_{\alpha}$, $|\Delta_{\alpha0}|$, and
$\gamma_{\alpha}$ for different $\alpha$ are close, the term $\mathcal{J}
_{\Delta,\mathrm{pm}}^{(2)}$ can be roughly estimated as
\end{subequations}
\begin{equation}
\mathcal{J}_{\Delta,\mathrm{pm}}^{(2)}\sim\pi T_{c}\frac{\Delta_{s0}^{2}}{\tilde{\gamma
}_{B1}\gamma_{B1}\Delta_{10}^{2}} \label{J2DpmEst}
\end{equation}
and we can see that this term is typically smaller than $\mathcal{J}
_{b,\mathrm{pm}}^{(2)}$, Eq.\ (\ref{J2bpmResult}), by the ratio
$\Delta_{s0}/\Delta_{10}$ (assuming $\Delta_{10}/\pi T_{c}\sim1$).

\subsection{Region near $T_{c}^{s}$ for $T_{c}^{s}\ll T_{c}$}

Near the transition temperature of the s-wave superconductor all terms contributing
to $\mathcal{J}^{(2)}$ decrease as $\Delta_{s0}^{2}$. In particular, the dominating
term $\mathcal{J}_{b,\mathrm{pm}}^{(2)}$ behaves as
\begin{equation}
\mathcal{J}_{b,\mathrm{pm}}^{(2)}\approx-\frac{\pi}{4}\sum_{\alpha}\sqrt
{\frac{\pi T_{c}}{|\Delta_{\alpha0}|}} \frac{\Delta_{s0}^{2}/T_{c}^{s}}
{\tilde{\gamma}_{B\alpha}\gamma_{B\alpha}} \text{ for } T\!\rightarrow\!
T_{c}^{s}\!-\!0.
\end{equation}
This behavior has an important consequence: the width of the TRSB state shrinks with
increasing temperature. On the other hand, the weak coupling approach breaks down
when the temperature is too close to $T_{c}^{s}$ when the correction to the $s$-wave
gap becomes comparable with its bulk value.

\section{Width of TRSB region \label{Sec:TRSBWidth}}

To analyze the width of the TRSB region in the weak-coupling regime, we represent
the supercurrent flowing through the interface in the form
$j=\sum_{\alpha}j_{\alpha}=\mathcal{A}_{0}\left(  \mathcal{J}^{(1)}\sin
\phi+\mathcal{J}^{(2)}\cos\phi\sin\phi\right)  $. The transition roughly corresponds
to the vanishing of the first harmonic $\mathcal{J}^{(1)}$ which we can represent as
\begin{equation}
\mathcal{J}^{(1)}  =\sum_{\alpha}\frac{(-1)^{\alpha}}{\tilde{\gamma
}_{B\alpha}}\Delta_{s0}f_{J,\alpha},
\label{J1pres}
\end{equation}
with
\begin{align*}
f_{J,\alpha}  &  \equiv f_{J}\!\left(  \frac{|\Delta_{\alpha0}
|}{T},\frac{\Delta_{s0}}{T}\right) \\
&  =\!2\pi T\!\sum_{\omega>0}\!\frac{|\Delta_{\alpha0}|}{\sqrt{\omega
^{2}\!+\!|\Delta_{\alpha0}|^{2}} \sqrt{\omega^{2}\!+\!|\Delta_{s0}|^{2}}}.
\end{align*}
We remind that the TRSB state only exists if $\mathcal{J}^{(2)}<0$. In this case,
which is realized for our system, we can write the condition for the TRSB state
range as
\begin{equation}
\left\vert \frac{f_{J,1}}{\tilde{\gamma}_{B1}}-\frac{f_{J,2} }{\tilde{\gamma
}_{B2}}\right\vert <\frac{|\mathcal{J}^{(2)}|}{\Delta_{s0}}.
\label{TRSBcondit}
\end{equation}
This formula together with microscopic results for $\mathcal{J}^{(2)}$ of the
previous section represent the main results of this paper. For fixed
$\tilde{\gamma}_{B1}^{-1}$ the transition from aligned to TRSB state occurs at the
following values of $\tilde{\gamma}_{B2}^{-1}$
\[
\frac{1}{\tilde{\gamma}_{B2}}=\frac{f_{J,1}}{f_{J,2}\tilde{\gamma}_{B1}}
\pm\frac{\mathcal{J}^{(2)}}{\Delta_{s0}f_{J,2}}.
\]
As $\mathcal{J}^{(2)}$ scales as $\tilde{\gamma}_{B\alpha}^{-2}$, The width of the
TRSB\ region can be conveniently characterized by the parameter
\begin{equation}
\tilde{\gamma}_{B2}^{2}\Delta\tilde{\gamma}_{B2}^{-1}\approx\Delta
\tilde{\gamma}_{B2}=\frac{2\tilde{\gamma}_{B2}^{2}\mathcal{J}^{(2)}}
{\Delta_{s0}f_{J,2}}, \label{TRSBwidth}
\end{equation}
which \emph{depends only on bulk properties of the superconductors} and does not
depend on the boundary resistances.\cite{Note39Bulk} In particular, at low
temperatures and for $\Delta_{s0}\ll|\Delta_{20}|$, using the asymptotic
$f_{J,\alpha}\approx \ln (4|\Delta_{\alpha0}|/\Delta_{s0})$ and keeping only the
dominating term in $\mathcal{J}^{(2)}$, Eq.\ (\ref{J2bpmResult}), we obtain the
following estimate
\begin{equation}
\Delta\tilde{\gamma}_{B2}\approx\pi\ln\!\frac{4|\Delta_{20}|}{\Delta_{s0}}
\sum_{\alpha}\frac{\gamma_{\alpha}}{\left(  \ln\frac{4|\Delta_{\alpha0}
|}{\Delta_{s0}}\right)  ^{2}}\sqrt{\frac{\pi T_{c}}{|\Delta_{\alpha0}|}}.
\end{equation}
We can see that this width only weakly depends on the value of the $s$-wave gap and
is mostly determined by the parameters $\gamma_{\alpha}$. Using the definition of
$\tilde{\gamma}_{B\alpha}$ and $\gamma_{\alpha}$, Eqs.\ (\ref{gBa}) and (\ref{ga}),
we immediately obtain a simple estimate for the spread of the partial boundary
resistance $\Delta R_{B2}$ within which the TRSB state exists,
\begin{equation}
\Delta R_{B2}\approx\pi\ln\!\frac{4|\Delta_{20}|}{\Delta_{s0}}
\sum_{\alpha}\frac{\rho_{\alpha} \xi_{\alpha}}{\left(  \ln\frac{4|\Delta_{\alpha0}
|}{\Delta_{s0}}\right)  ^{2}}\sqrt{\frac{\pi T_{c}}{|\Delta_{\alpha0}|}},
\end{equation}
which is mainly determined by the products $\rho_{\alpha} \xi_{\alpha}$.

\begin{figure}[ptb]
\includegraphics[width=3.4in]{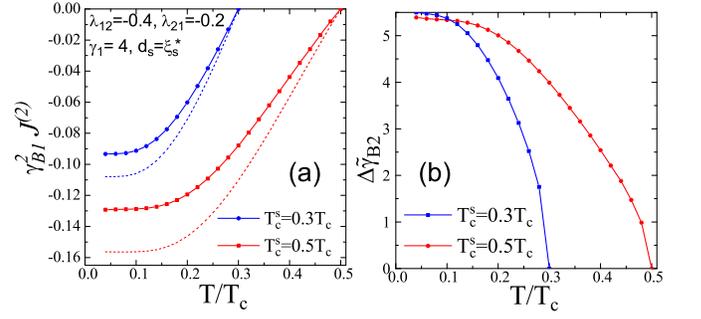}
\caption{(a)Temperature
dependences of the second-harmonic amplitude for two values of the ratio
$T_{c}^{s}/T_{c}$, 0.3 and 0.5. Dashed lines show the term $\mathcal{J}
_{b,\mathrm{pm}}^{(2)}$ only. Other parameters are shown in the plot. We also
assume $\lambda_{11}\!=\!\lambda_{22}\!=\!0$. (b) Corresponding temperature dependences
of the TRSB width, Eq.\ (\ref{TRSBwidth}).}
\label{Fig-J2Tdep}
\end{figure}

Figure \ref{Fig-J2Tdep} illustrates the temperature dependences of the
second-harmonic amplitude $\mathcal{J}^{(2)}$ and the width of the TRSB region for
the representative parameters listed in the left figure and for the two values of
the ratio $T_{c}^{s}/T_{c}$, 0.3 and 0.5. In Fig. \ref{Fig-J2Tdep}(a) we show for
comparison both the full amplitude of the second harmonic and the term
$\mathcal{J}_{b,\mathrm{pm}}^{(2)}$ only (dashed lines). We can see that this term
typically accounts for $80\%-85\%$ of the total amplitude. The remaining part mostly
comes from the terms $\mathcal{J} _{\Delta,\mathrm{pm}}^{(2)}$ and
$\mathcal{J}_{\phi,0}^{(2)}$. For used representative parameters the contributions
from the $s$-wave terms $\mathcal{J}_{\ast,s}^{(2)}$ are negligible. We emphasize
the shrinking of the TRSB width illustrated in Fig.\ \ref{Fig-J2Tdep}(b),
$\Delta\tilde{\gamma}_{B2} \propto\sqrt{T_{c}^{s}-T}$ near $T_{c}^{s}$. This means
that in some range of parameters, the transition from the aligned to TRSB state may
be observed as a function of temperature, as in the case of the $0$-$\pi$ transition
in SFS junctions.\cite{RyazSFS0Pi}

\section{Density of states of s-wave superconductor within the TRSB region
\label{Sec:DoSTRSB}}

\begin{figure}[ptb]
\includegraphics[width=0.4\textwidth]{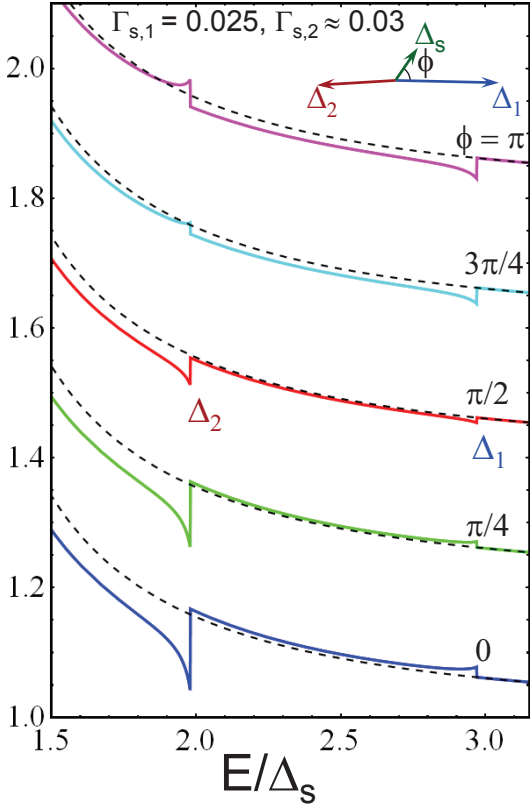}\caption{ Evolution of
the proximity-induced features in the $s$-wave DoS within the TRSB region with
increasing angle $\phi$ between $\Delta_{s0}$ and $\Delta_{10}$. Curves are
vertically displaced for clarity. Dashed lines show the bulk DoS.}
\label{Fig-DoS-TRSB}
\end{figure}

Contact with $s_{\pm}$ superconductor induces specific features in the  $s$-wave
density of states (DOS).  For the \emph{aligned} states in the case of thin $s$-wave
layer, the $s_{\pm}$ gaps aligned with $\Delta_{s0}$ generate positive corrections
to DoS, while anti-aligned gaps generate negative
corrections.\cite{ProxFingeprintEPL11} The latter negative features can be used to
identify $s_{\pm}$ state.

In this short section we consider evolution of contact-induced features of the
$s$-wave DoS across the TRSB region. To find the density of states, we have to
perform analytical continuation of the Green's functions to real energies
$i\omega\rightarrow E+i\delta$. The normalized DoS is related to the real-energy
Green's function by the standard expression,
\begin{equation}
N(E,x)=\mathrm{Re}\left[  \frac{E}{\sqrt{E^{2}-\Phi(E,x)\Phi^{\dagger}
(E,x)}}\right],
\end{equation}
where $\Phi^{\dagger}(E,x)=\Phi^{\ast}(-E,x)$. Expanding $\Phi_{s}(E,x)$ and taking
into account that $\Delta_{s0}$ is selected real, we obtain the proximity-induced
correction to the $s$-wave DoS
\begin{equation}
\delta N_{s}(E,x)\approx\mathrm{Re}\left[  \frac{E\Delta_{s0}\left(
\tilde{\Phi}_{s}^{\dagger}(E,x)+\tilde{\Phi}_{s}(E,x)\right)  }{2\left(
E^{2}-\Delta_{s0}^{2}\right)  ^{3/2}}\right]  \label{DoSCorrFormula}
\end{equation}

The correction to the Green's function can be represented again as a Fourier series,
$\tilde{\Phi}_{s}(E,x)=\sum_{m=0}^{\infty}\tilde{\Phi}_{s,m}(E)\cos\left( m\pi
x/d_{s}\right)$. In the Matsubara presentation, two contributions to the Fourier
components $\tilde{\Phi}_{s,m}$ are given by Eqs.\ (\ref{Phisbm}) and
(\ref{PhisDm-Dm}). Using these results, we obtain for these contributions at real
energies,
\begin{align*}
\frac{\tilde{\Phi}_{s,b,m}\!+\!\tilde{\Phi}_{s,b,m}^{\dagger}}{2}\! &
=\!\frac{(2\!-\!\delta_{m})\xi_{s,\Delta}^{2}/(d_{s}\xi_{s}^{\ast})}
{\sqrt{E^{2}\!-\!\Delta_{s0}^{2}}/\Delta_{s0}\!+\!i\left(  \pi m\xi_{s,\Delta
}/d_{s}\right)  ^{2}}\\
\times&  \sum_{\alpha}\frac{\sqrt{E^{2}\!-\!\Delta_{s0}^{2}}}{\sqrt
{|\Delta_{\alpha0}|^{2}\!-\!E^{2}}}\frac{\operatorname{Re}[\Delta_{\alpha
0}]\!-\!\Delta_{s0}}{\tilde{\gamma}_{B\alpha}},\\
\frac{\tilde{\Phi}_{s,\Delta,m}\!+\!\tilde{\Phi}_{s,\Delta,m}^{\dagger}}{2} &
=\!\frac{\sqrt{E^{2}\!-\!\Delta_{s0}^{2}}\tilde{\Delta}_{s,m}^{R}}{\sqrt
{E^{2}\!-\!\Delta_{s0}^{2}}+i\Delta_{s0}\left(  \pi m\xi_{s,\Delta}
/d_{s}\right)  ^{2}},
\end{align*}
where $\tilde{\Delta}_{s,m}^{R}$ is given by Eq.\ (\ref{CorDeltasTxt}a). These
results together with Eq.\ (\ref{DoSCorrFormula}) determine the shape of the DoS
correction for arbitrary parameters of superconductors in the linear approximation
with respect to the coupling strength $1/\tilde{\gamma}_{B\alpha}$.

For the important case of small $d_{s}$ we can keep only the uniform $m\!=\!0$ term
in the Fourier expansions leading to a simple result similar to one for the aligned
state\cite{ProxFingeprintEPL11},
\begin{equation}
\frac{\tilde{\Phi}_{s,b,m}+\tilde{\Phi}_{s,b,m}^{\dagger}}{2}\!=\!\sum
_{\alpha}\Gamma_{B,\alpha}\frac{\operatorname{Re}[\Delta_{\alpha0}
]\!-\!\Delta_{s0}}{\sqrt{|\Delta_{\alpha0}|^{2}\!-\!E^{2}}}
\end{equation}
with
\[
\Gamma_{B,\alpha}=\Delta_{s0}\frac{\xi_{s,\Delta}^{2}}{d_{s}\xi_{s\tilde
{\gamma}_{B\alpha}}^{\ast}}=\frac{1}{2e^{2}\nu_{s}d_{s}R_{B}^{\alpha}}.
\]
Therefore, the correction to the s-wave DoS in the case of thin $s$-wave layer is
given by,
\begin{align}
\delta N_{s}(E,x)&\approx\frac{E\Delta_{s0}}{\left(  E^{2}-\Delta_{s0}
^{2}\right)  ^{3/2}}\sum_{\alpha}\Gamma_{B\alpha}\nonumber\\
\times&\frac{\operatorname{Re}[\Delta_{\alpha0}]\!-\!\Delta_{s0}}
{\sqrt{|\Delta_{\alpha0}|^{2}\!-\!E^{2}}}\Theta(|\Delta_{\alpha0}|-E),
\label{DoSResultresult}
\end{align}
where $\Theta(x)$ is the step function. If, as before, we define that phase shift
between $\Delta_{s0}$ and $\Delta_{10}$ as $\phi$ then $\operatorname{Re}
[\Delta_{10}]=|\Delta_{10}|\cos \phi$ and $\operatorname{Re}
[\Delta_{20}]=-|\Delta_{20}|\cos \phi$.

Figure \ref{Fig-DoS-TRSB} illustrates the evolution of this correction with
increasing angle $\phi$ for representative parameters. Two limiting cases $\phi=0$
and $\pi$ correspond to aligned states in which the aligned and anti-aligned gaps
induce asymmetric peak and dip correspondingly. With increasing phase, the peak
smoothly transforms into a dip and vice versa. In the maximally frustrated state for
$\phi=\pi/2$ the DoS has two small dips.  Note that the  DoS correction is obtained
with the linear approximation with respect to the coupling between the
superconductors, which somewhat overestimates the amplitude and sharpness of the
peaks.\cite{ProxFingeprintEPL11}

\section{Summary and discussion}

In summary, we evaluated the range of parameters where the TRSB state is realized
for the interface between $s$-wave and $s_{\pm}$ superconductors using the simple
microscopic theory. This state appears when the partial Josephson energies almost
completely compensate each other. The width of the TRSB region is determined by the
$\sin (2\phi)$ term in the Josephson current, which appears in the second order with
respect to the coupling strength between the superconductors. We found that the
dominating contribution to this term is determined by the direct boundary correction
to the Green's function of the $s_{\pm}$ superconductor. This term is missed by
phenomenological models of the junction. The width of the TRSB region shrinks with
increasing temperature giving possibility to detect the transition from the aligned
to TRSB state as function of temperature.

The main purpose of this paper is to establish factors which determine the width of
the TRSB region at the $s$/$s_{\pm}$ interface in the simplest possible situation
accessible for full analytical analysis. Even in this relatively simple case the
analysis occurred to be very nontrivial.

Several factors may have quantitative influence on results reported in this paper
and complicate their applications to real iron-based superconductors:
\begin{itemize}
  \item
  Most of these materials have more than two bands (up to five). This is not a
  crucial complication. Our consideration can be directly generalized to
  arbitrary number of bands.
  \item Due to the
  very short coherence length, most iron-based superconductors are probably in
  the clean limit. In the cleanest materials a significant anisotropy of the gap
  \cite{GapAnis} and even the gap nodes\cite{nodes} were revealed. On the other
  hand, in several other compounds isotropic gaps within the bands were
  found\cite{IsGaps}, which probably indicates substantial intraband scattering.
  The presence of a significant gap anisotropy and nodes does not contradict an
  overall picture of the $s_{\pm}$ state because what matters most is the
  average gap inside the band. This means that the TRSB state is also expected
  within some range of parameters at the interface between conventional and
  clean $s_{\pm}$ superconductor. However, our calculation of the width of this
  region is not directly applicable to this case.
  \item In compounds with strong impurity scattering one can expect some interband
  scattering, which was neglected in our model. As this scattering suppresses
  $s_{\pm}$ state, it can not be too strong. The main effects of this scattering
  are suppression of the $s_{\pm}$ gap parameters and emergence of the subgap
  states. These effect may have some influence on the location and width of the
  TRSB region.
\end{itemize}
An accurate description of these factors requires special considerations that will
further complicate the theoretical model. Nevertheless, we expect that a qualitative
picture of the transitional region will hold within a more realistic framework.

\begin{acknowledgments}
I acknowledge many useful discussions with Valentin Stanev and Thomas Proslier. This
work was supported by UChicago Argonne, LLC, operator of Argonne National
Laboratory, a U.S. Department of Energy Office of Science laboratory, operated under
contract No. DE-AC02-06CH11357, and by the \textquotedblleft Center for Emergent
Superconductivity\textquotedblright, an Energy Frontier Research Center funded by
the U.S. Department of Energy, Office of Science, Office of Basic Energy Sciences
under Award Number DE-AC0298CH1088.
\end{acknowledgments}

\appendix

\section{Derivation of boundary-induced corrections in the weak-coupling limit
\label{App-WeakCoupl}}

In this Appendix we derive corrections to the Green's functions and gaps in the
linear order with respect to the coupling parameters $1/\gamma_{B\alpha}$. As a zero
approximation, we consider a general complex case when there is a finite phase
difference $\phi$ between the bulk zero-order gap parameters $\Delta_{10}$ and
$\Delta_{s0}$ as shown in Fig.\ \ref{Fig-DeltasFrustrState}. Correspondingly, the
phase difference between $\Delta_{20}$ and $\Delta_{s0}$ is $\phi-\pi$. This
calculation covers both aligned states when $\phi$ equals $0$ or $\pi$ and complex
TRSB states with $0<\phi<\pi$.

\subsection{$s$-wave gap and Green's function \label{App-weak-s}}

We start with calculation of corrections to the $s$-wave Green's functions and gap
parameter, $\tilde{\Phi}_{s}$ and $\tilde{\Delta}_{s}$.  From Eq.\ (\ref{Ueqs1}) we
obtain that the first-order corrections to the $s$-wave Green's functions obeys the
following equations
\begin{equation}
\xi_{s,\omega}^{2}\frac{d^{2}\tilde{\Phi}_{s}}{dx^{2}}-\tilde{\Phi}_{s}
=-\tilde{\Delta}_{s}, \label{UeqLin}
\end{equation}
where $\xi_{s,\omega}^{2}\!
=\!D_{s}/\left(2\sqrt{\omega^{2}\!+\!\Delta_{s0}^{2}}\right)\!=\!\xi_{s,\Delta}^{2}
\Delta_{s0}/\sqrt{\omega^{2}\!+\!\Delta_{s0}^{2}}$ and $\xi_{s,\Delta}
^{2}=D_{s}/(2\Delta_{s0})$. Without loss of generality, the zero-order gap
parameters $\Delta_{s0}$ can be selected real. In this case the self-consistency
condition for the linear corrections can be written as
\begin{equation}
2\pi T\sum_{\omega>0}\frac{1}{\sqrt{\omega^{2}\!+\!\Delta_{s0}^{2}}}\left(
\tilde{\Phi}_{s}\!-\!\frac{\Delta_{s0}^{2}\operatorname{Re}[\tilde{\Phi}_{s}
]}{\omega^{2}\!+\!\Delta_{s0}^{2}}\!-\!\tilde{\Delta}_{s}\right) \! =\!0.
\label{SelfConsLins}
\end{equation}
In the boundary condition for $d\Phi_{s}/dx$, Eq.\ (\ref{BCgen1}), we can neglect in
the right hand side differences between $\Phi$'s and $\Delta$'s and approximate
$\Delta$'s by their bulk values. This gives
\begin{equation}
\frac{\xi_{s}^{\ast}}{\sqrt{\omega^{2}+\Delta_{s0}^{2}}}\frac{d\tilde{\Phi}
_{s}}{dx}=-\sum_{\alpha}\frac{1}{\tilde{\gamma}_{B\alpha}}\frac
{\Delta_{s0}-\Delta_{\alpha0}}{\sqrt{\omega^{2}+|\Delta_{\alpha0}|^{2}}}
\label{BoundCondLins}
\end{equation}
at $x=0$. Note that, in general, $\tilde{\Delta}_{s}$, $\tilde{\Phi}_{s}$, and
$\Delta_{\alpha0}$ are complex, $\Delta_{\alpha0}=|\Delta_{\alpha0}|\exp
(i\phi_{\alpha})$, $\phi_{1}=\phi$, $\phi_{2}=\phi-\pi$. They are real only for the
aligned state.

To solve Eqs. (\ref{UeqLin}) and (\ref{BoundCondLins}) it is convenient to split
$\tilde{\Phi}_{s}$ into the two contributions, $\tilde{\Phi}_{s}
=\tilde{\Phi}_{s,b}+\tilde{\Phi}_{s,\Delta}$, where $\tilde{\Phi}_{s,b}$ is induced
by the boundary condition and $\tilde{\Phi}_{s,\Delta}$ is induced by the gap
correction. The first contribution $\tilde{\Phi}_{s,b}$ can be found from the
following equation and the boundary condition
\begin{subequations}
\begin{align}
&  \xi_{s,\omega}^{2}\tilde{\Phi}_{s,b}^{\prime\prime}-\tilde{\Phi}
_{s,b}=0,\label{EqPhi_sb}\\
&  \xi_{s}^{\ast}\tilde{\Phi}_{s,b}^{\prime}(0)\!=\!\sum_{\alpha}\frac{1}
{\tilde{\gamma}_{B\alpha}}\frac{\sqrt{\omega^{2}\!+\!\Delta_{s0}^{2}}}
{\sqrt{\omega^{2}\!+\!|\Delta_{\alpha0}|^{2}}}\left(\Delta_{\alpha0}\!-\!\Delta
_{s0}\right)  , \label{BCPhi_sb0}
\end{align}
while the second contribution, $\tilde{\Phi}_{s,\Delta}$, obeys
\end{subequations}
\begin{subequations}
\begin{align}
\xi_{s,\omega}^{2}\tilde{\Phi}_{s,\Delta}^{\prime\prime}-\tilde{\Phi
}_{s,\Delta}  &  =-\tilde{\Delta}_{s},\label{EqPhiD}\\
\tilde{\Phi}_{s,\Delta}^{\prime}(0)  &  =0. \label{BCPhiD}
\end{align}
The solution $\tilde{\Phi}_{s,b}(x)$ of the linear equation (\ref{EqPhi_sb}) with
the boundary condition $\tilde{\Phi}_{s,b}^{\prime}=0$ at $x=-d_{s}$ is given by
\end{subequations}
\begin{equation}
\tilde{\Phi}_{s,b}(x)=C_{s,b}\cosh\left(  \frac{x+d_{s}}{\xi_{s,\omega}
}\right)  , \label{Phi-sbSol}
\end{equation}
where the constant $C_{s,b}$ can be found from the boundary condition at $x=0$, Eq.\
(\ref{BCPhi_sb0}),
\[
C_{s,b}=\frac{\xi_{s,\omega}/\xi_{s}^{\ast}}{\sinh\left(  d_{s}/\xi_{s,\omega
}\right)  }\sum_{\alpha}\frac{\sqrt{\omega^{2}+\Delta_{s0}^{2}}}{\tilde
{\gamma}_{B\alpha}\sqrt{\omega^{2}+|\Delta_{\alpha0}|^{2}}}(\Delta_{\alpha
0}-\Delta_{s0})
\]
leading to the following result
\begin{align}
\tilde{\Phi}_{s,b}(x)  &  =\frac{\xi_{s,\omega}\cosh\left[  \left(
x+d_{s}\right)  /\xi_{s,\omega}\right]  }{\xi_{s}^{\ast}\sinh\left(  d_{s}
/\xi_{s,\omega}\right)  }\nonumber\\
&  \times\sum_{\alpha}\frac{\sqrt{\omega^{2}+\Delta_{s0}^{2}}}{\tilde{\gamma
}_{B\alpha}\sqrt{\omega^{2}+|\Delta_{\alpha0}|^{2}}}(\Delta_{\alpha0}
-\Delta_{s0}). \label{Phis-b-x}
\end{align}

We compute $\tilde{\Phi}_{s,\Delta}$ and $\tilde{\Delta}_{s}$ using the Fourier
expansions, $\tilde{\Phi}_{s,\Delta}(x)\!=\!\sum_{m=0}^{\infty}\tilde{\Phi
}_{s,\Delta,m}\cos k_{m}x$,$\ \tilde{\Delta}_{s}(x)\!=\!\sum_{m=0}^{\infty}
\tilde{\Delta}_{s,m}\cos k_{m}x$ with $k_{m}=m\pi/d_{s}$. The Fourier components
$\tilde{\Phi}_{s,b,m}$ of $\tilde{\Phi}_{s,b}(x)$ can be computed explicitly from
Eq.\ (\ref{Phis-b-x}),
\begin{equation}
\tilde{\Phi}_{s,b,m}\!=\!\frac{(2\!-\!\delta_{m})\xi_{s,\omega}^{2}/(d_{s}
\xi_{s}^{\ast})}{1+\xi_{s,\omega}^{2}k_{m}^{2}}\sum_{\alpha}\frac{\sqrt
{\omega^{2}\!+\!\Delta_{s0}^{2}}}{\sqrt{\omega^{2}\!+\!|\Delta_{\alpha0}|^{2}
}}\frac{\Delta_{\alpha0}\!-\!\Delta_{s0}}{\tilde{\gamma}_{B\alpha}}.
\label{Phisbm}
\end{equation}
Eq. (\ref{EqPhiD}) immediately gives the following relation between the Fourier
components $\tilde{\Phi}_{s,\Delta,m}$ and $\tilde{\Delta}_{s,m}$
\begin{equation}
\tilde{\Phi}_{s,\Delta,m}=\frac{\tilde{\Delta}_{s,m}}{1+\xi_{s,\omega}
^{2}k_{m}^{2}}. \label{PhisDm-Dm}
\end{equation}
Substituting this result into the self-consistency condition (\ref{SelfConsLins})
and splitting it into the real and imaginary parts, we relate
$\tilde{\Delta}_{s,m}=\tilde{\Delta}_{s,m}^{R}+i\tilde{\Delta} _{s,m}^{I}$ to
$\tilde{\Phi}_{s,b,m}=\tilde{\Phi}_{s,b,m}^{R}+i\tilde{\Phi }_{s,b,m}^{I}$ as
\begin{subequations}
\label{DmPhisbm}
\begin{align}
&  2\pi T\sum_{\omega>0}\frac{\Delta_{s0}^{2}\!+\!\omega^{2}\frac
{\xi_{s,\omega}^{2}k_{m}^{2}}{1+\xi_{s,\omega}^{2}k_{m}^{2}}}{\left(
\omega^{2}+\Delta_{s0}^{2}\right)  ^{3/2}}\tilde{\Delta}_{s,m}^{R}\!=\!2\pi
T\sum_{\omega>0}\frac{\omega^{2}\tilde{\Phi}_{s,b,m}^{R}}{\left(  \omega
^{2}+\Delta_{s0}^{2}\right)  ^{3/2}},\label{DmPhisbmR}\\
&  2\pi T\sum_{\omega>0}\frac{1}{\sqrt{\omega^{2}\!+\!\Delta_{s0}^{2}}}
\frac{\xi_{s,\omega}^{2}k_{m}^{2}}{1\!+\!\xi_{s,\omega}^{2}k_{m}^{2}}
\tilde{\Delta}_{s,m}^{I}\!=\!2\pi T\sum_{\omega>0}\!\frac{\tilde{\Phi}
_{s,b,m}^{I}}{\sqrt{\omega^{2}\!+\!\Delta_{s0}^{2}}}. \label{DmPhisbmI}
\end{align}
The components $\tilde{\Delta}_{s,m}^{R}$ and $\tilde{\Delta}_{s,m}^{I}$ describe
responses of the s-wave gap parameter to the interface perturbation in the amplitude
and phase channels. Eqs. (\ref{Phisbm}), (\ref{PhisDm-Dm}), and (\ref{DmPhisbm})
already provide a formal solution of the problem. As the left hand side of Eq.\
(\ref{DmPhisbmI}) vanishes for $m\!=\!0$, solution for the imaginary part exists
only if
\end{subequations}
\[
\sum_{\omega>0}\frac{\tilde{\Phi}_{s,b,0}^{I}}{\sqrt{\omega^{2}+\Delta
_{s0}^{2}}}=0
\]
giving the condition
\begin{equation}
\sum_{\alpha,\omega>0}\frac{1}{\sqrt{\omega^{2}+\Delta_{s0}^{2}}\sqrt
{\omega^{2}+|\Delta_{\alpha0}|^{2}}}\frac{\operatorname{Im}[\Delta_{\alpha0}
]}{\tilde{\gamma}_{B\alpha}}=0. \label{ImCond}
\end{equation}
Uncertainty in $\tilde{\Delta}_{s,0}^{I}$ reflects the phase-rotation invariance and
we can select $\tilde{\Delta}_{s,0}^{I}=0$. As the Josephson energy between the
$s$-wave superconductor and $\alpha$ band, $E_{J\alpha}$ is given by Eq.\
(\ref{PartJosEner}) and $\Delta_{\alpha0}=|\Delta_{\alpha0}|\exp(i\phi_{\alpha})$,
we can see that the condition (\ref{ImCond}) simply means
\[
\sum_{\alpha}E_{J\alpha}\sin\phi_{\alpha}=0.
\]
Since the partial Josephson currents are proportional to $E_{J\alpha}$, this
condition implies that the total Josephson current flowing through the boundary is
always zero in the ground state. For two bands the condition for realization of the
TRSB state \emph{in the linear order with respect to the interface transparency} is
simply $E_{J1}=E_{J2}$. To establish an accurate range of parameters within which
the TRSB state is stable, one has to go beyond the linear order.

Using the expansion (\ref{Phisbm}), we obtain the explicit presentations for
$\tilde{\Delta}_{s,m}^{R}$ and $\tilde{\Delta}_{s,m}^{I}$ given by Eqs.\
(\ref{CorDeltasTxt}) of the main text which in turn determine the term
$\tilde{\Phi}_{s,\Delta,m}$ of the Green's function, Eq.\ (\ref{PhisDm-Dm}). Since
we already derived the result for $\tilde{\Phi}_{s,b,m}$, Eq.\ (\ref{Phis-b-x}), we
now have all corrections.

The full analytical formulas are somewhat cumbersome and it is useful to derive more
transparent results in simple limiting cases. At low temperatures the summation with
respect to the Matsubara frequencies can be replaced by the integration $2\pi
T\sum_{\omega>0}\rightarrow\int_{0}^{\infty}d\omega$. In this limit we can obtain
the analytical result for the average correction to the order parameter amplitude,
$\tilde{\Delta}_{s,0}^{R}$,
\begin{equation}
\frac{\tilde{\Delta}_{s,0}^{R}}{\pi T_{c}}  =\frac{\xi_{s}^{\ast}}{d_{s}
}\sum_{\alpha}U(\Delta_{s0}/|\Delta_{\alpha0}|)\frac{\operatorname{Re}
[\Delta_{\alpha0}]-\Delta_{s0}}{\tilde{\gamma}_{B\alpha}|\Delta_{\alpha0}
|}\label{AvDeltasLowTApp}
\end{equation}
with
\[
U(a) \! =\!
\int\limits_{0}^{\infty }\!dz\frac{z^{2}}{(z^{2}\!+\!1) ^{3/2}\sqrt{a^{2}z^{2}\!+\!1}}
 \! =\!\frac{K(1\!-\!a^{2})-E(1\!-\!a^{2})}{1-a^{2}},
\]
where $K(m)\!=\!\int_{0}^{\pi/2}(1\!-\!m\sin^{2}\theta)^{-1/2}d\theta$ and
$E(m)\!=\!\int_{0}^{\pi/2}(1\!-\!m\sin^{2}\theta)^{1/2}d\theta$ are the complete
elliptic integrals. As we mentioned before, the uniform part of $\tilde
{\Delta}_{s}$ can be selected to be real, $\tilde{\Delta}_{s,0}=\tilde{\Delta
}_{s,0}^{R}$.

We can derive the simple analytical results for important particular case of (i)
thin $s$-layer, $d_{s}\ll\xi_{s,\Delta}$, (ii) weaker $s$-superconductor,
$\Delta_{s0}\ll|\Delta_{\alpha0}|$, and (iii) low temperatures, $T\ll T_{c}^{s}$.
Due to the first condition, the dominating contribution to the gap correction is
given by the coordinate independent part $\tilde{\Delta}_{s,0}$, which is determined
by the general formula (\ref{AvDeltasLowTApp}). In the limit of
$\Delta_{s0}\ll|\Delta_{\alpha0}|$ we can use the asymptotics of the function $U(a)$
in the limit $a\ll1$, $U(a)\approx\ln\left(  4/a\right)  -1$, leading to the
following simple result
\begin{equation}
\frac{\tilde{\Delta}_{s,0}^{R}}{\pi T_{c}}\approx\!\frac{\xi_{s}^{\ast}}{d_{s}
}\sum_{\alpha}\!\frac{\operatorname{Re}[\Delta_{\alpha0}]\!-\!\Delta_{s0}}
{\tilde{\gamma}_{B\alpha}|\Delta_{\alpha0}|}\left[  \ln\!\left(  \frac
{4|\Delta_{\alpha0}|}{\Delta_{s0}}\right) \! -\! 1\right]  . \label{DsLimCaseApp}
\end{equation}
The sign of $\tilde{\Delta}_{s,0}^{R}$ determines net effect of the $s_{\pm}$
superconductor on the $s$ superconductor, i.e., the sign of the proximity effect
(positive vs negative proximity). The proximity is always negative in the TRSB
state.  In the case of the aligned state corresponding to
$\operatorname{Re}[\Delta_{\alpha0}]=\Delta_{\alpha0}$, as one can expect, $s_{\pm}$
gaps aligned with $\Delta_{s0}$ enhance $s$-wave superconductivity while
anti-aligned $s_{\pm}$ gaps suppress superconductivity in the $s$ superconductor.
The relative contributions are mostly determined by the electrical coupling between
$s$-superconductor and the $s_{\pm}$ bands described by parameters
$\tilde{\gamma}_{B\alpha}^{-1}$. Another important factor is that the aligned gaps
give positive contribution proportional to gap difference
$\Delta_{\alpha0}-\Delta_{s0}$, while antialigned gaps give negative contribution
proportional to gap sum $|\Delta_{\alpha0}|+\Delta_{s0} $. This give possibility to
total negative proximity effect in the aligned state.

Weak spatial dependence of $\tilde{\Delta}_{s}(x)$ is determined by the components
$\tilde{\Delta}_{s,m}$ with $m>0$. At $T=0$ these components can be presented as
\begin{subequations}
\begin{align}
\tilde{\Delta}_{s,m}^{R}  & \! =\!\frac{1}{Z_{s,m}^{a}}\frac{2\xi_{s,\Delta}^{2}
}{d_{s}\xi_{s}^{\ast}}\sum_{\alpha}\!J_{a}\!\left( \frac{|\Delta_{\alpha0}
|}{\Delta_{s0}},\beta_{m}\right)  \frac{\operatorname{Re}[\Delta_{\alpha
0}]\!-\!\Delta_{s0}}{\tilde{\gamma}_{B\alpha}},\label{DsxLimCaseGenR}\\
\tilde{\Delta}_{s,m}^{I}  &  \!=\!\frac{1}{Z_{s,m}^{\phi}}\frac{2\xi_{s,\Delta
}^{2}}{d_{s}\xi_{s}^{\ast}}\sum_{\alpha}\!J_{\phi}\!\left( \frac{|\Delta
_{\alpha0}|}{\Delta_{s0}},\beta_{m}\right)  \frac{\operatorname{Im}
[\Delta_{\alpha0}]}{\tilde{\gamma}_{B\alpha}} \label{DsxLimCaseGenI}
\end{align}
with $\beta_{m}=\left(  \pi m\xi_{s,\Delta}/d_{s}\right)  ^{2}$,
\end{subequations}
\begin{align*}
Z_{s,m}^{a}  &  =\int_{0}^{\infty}\frac{dz}{(z^{2}+1)^{3/2}}\left[
1+\frac{\beta_{m}z^{2}}{\sqrt{z^{2}+1}+\beta_{m}}\right] \\
&  =\frac{1}{\beta_{m}}\left[  \frac{\pi}{2}+\sqrt{\beta_{m}^{2}-1}\ln\left(
\sqrt{\beta_{m}^{2}-1}+\beta_{m}\right)  \right]  ,\\
J_{a}(\delta,\beta)  &  =\int_{0}^{\infty}dz\frac{z^{2}}{\left(
z^{2}+1\right)  \sqrt{z^{2}+\delta^{2}}}\frac{1}{\sqrt{z^{2}+1}+\beta},\\
Z_{s,m}^{\phi}  &  =\int_{0}^{\infty}\frac{dz}{\sqrt{z^{2}+1}}\frac{\beta_{m}
}{\sqrt{z^{2}+1}+\beta_{m}}\\
&  =\frac{\beta_{m}}{\sqrt{\beta_{m}^{2}-1}}\ln\left(  \beta_{m}+\sqrt
{\beta_{m}^{2}-1}\right)  ,\\
J_{\phi}(\delta,\beta)  &  =\int_{0}^{\infty}\frac{dz}{\sqrt{z^{2}+\delta^{2}
}\left(  \sqrt{z^{2}+1}+\beta\right)  }.
\end{align*}
In the limit $\beta _{m}\gg1$ corresponding to $d_s\ll \xi_s$, asymptotics of both
$Z_{s,m}^{a}$ and $Z_{s,m}^{\phi}$ are $Z_{s,m}^{\{a,\phi\}}\approx\ln\left(
2\beta_{m}\right)  $. In the limits $|\Delta_{\alpha0}|\gg\Delta_{s0}$ and
$\Delta_{s0}\beta_{m} \gg|\Delta_{\alpha0}|$ corresponding to $\beta\gg\delta\gg1$,
$J_{a} (\delta,\beta)$ and $J_{\phi}(\delta,\beta)$ also have the same asymptotics
\[
J_{\{a,\phi\}}(\delta,\beta)\approx\int_{0}^{\infty}\frac{dz}{\left(
z+\beta\right)  \sqrt{z^{2}+\delta^{2}}}\approx\frac{1}{\beta}\ln\frac{2\beta
}{\delta}.
\]
Collecting all terms, we obtain
\[
\tilde{\Delta}_{s,m}\!\approx\!\sum_{\alpha}\!\left[ 1\!+\!\frac{\ln\left(  \Delta
_{s0}/|\Delta_{\alpha0}|\right)  }{\ln\left[  2\left(  \pi m\xi_{s,\Delta
}/d_{s}\right)  ^{2}\right]  }\right]  \frac{2d_{s}/\xi_{s}^{\ast}}{\left(
\pi m\right)  ^{2}}\frac{\Delta_{\alpha0}\!-\!\Delta_{s0}}{\tilde{\gamma}
_{B\alpha}}.
\]
Using the relation $(|x|\!-\!1)^{2}\!=\!\frac{1}{3}\!+\!4\sum_{m=1}^{\infty}
\frac{\cos(\pi mx)}{\left(  \pi m\right)^{2}}$, we can approximately present the gap
correction in real space as
\begin{align}
\tilde{\Delta}_{s}(x)\!  &  \approx\!\tilde{\Delta}_{s,0}\!-\!\frac{d_{s}}
{\xi_{s}^{\ast}}\!\sum_{\alpha}\frac{\Delta_{\alpha0}\!-\!\Delta_{s0}}
{\tilde{\gamma}_{B\alpha}}\left[  \frac{(x\!+\!d_{s})^{2}}{2d_{s}^{2}
}\!-\!\frac{1}{6}\right] \nonumber\\
&  \times\left[  1+\frac{\ln\left(  \Delta_{s0}/|\Delta_{\alpha0}|\right)
}{2\ln\left(  \pi\sqrt{2}\xi_{s,\Delta}/d_{s}\right)  }\right]  .
\label{DeltasRealLim}
\end{align}
Correspondingly, for the Green's function in the same limits we derive
\begin{align}
\tilde{\Phi}_{s}(\omega,x)  &  \approx\tilde{\Delta}_{s,0}+\pi T_{c}\frac
{\xi_{s}^{\ast}}{d_{s}}\left[  1+\frac{1}{2}\left(  \frac{x+d_{s}}
{\xi_{s,\omega}}\right)  ^{2}\right] \nonumber\\
&  \times\sum_{\alpha}\frac{1}{\sqrt{\omega^{2}+\Delta_{\alpha}^{2}}}
\frac{\Delta_{\alpha0}-\Delta_{s0}}{\tilde{\gamma}_{B\alpha}}. \label{PhisLim}
\end{align}
In summary, simple analytical results given by Eqs. (\ref{DsLimCaseApp}),
(\ref{DeltasRealLim}), and (\ref{PhisLim}) determine corrections to the $s$-wave gap
and Green's function for a thin $s$-wave layer.

\subsection{$s_{\pm}$-wave gaps and Green's functions \label{App-Weak-spm}}

We can evaluate corrections to the $s_{\pm}$ gap parameters and Green's functions
following the same general route. The difference is that the matrix structure of the
self-consistency condition, Eq.\ (\ref{Ueqspm2}), has to be properly accounted for.
The first-order correction to $\Phi_{\alpha}$ with respect to the coupling strength
$\gamma_{B\alpha}^{-1}$ is determined by the following equation and boundary
conditions,
\begin{subequations}
\begin{align}
&\xi_{\alpha,\omega}^{2}\tilde{\Phi}_{\alpha}^{\prime\prime}-\tilde{\Phi
}_{\alpha}  =-\tilde{\Delta}_{\alpha},\label{UeqPMLin}\\
&\xi_{\alpha}G_{\alpha}\tilde{\Phi}_{\alpha}^{\prime}=\frac{G_{s}}
{\gamma_{B\alpha}}(\Delta_{s0}-\Delta_{\alpha0}),     \text{ at
}x=0 \label{BoundConPMLin}
\end{align}
and $\Phi_{\alpha}^{\prime}\!=\!0$ at $x\!=\!d_{\mathrm{\pm}}$ with $G_{\alpha}
\approx\omega/\sqrt{\omega^{2}\!+\!|\Delta_{\alpha0}|^{2}}$, $\xi_{\alpha,\omega
}^{2}= D_{\alpha}/(2\sqrt{\omega^{2}+|\Delta_{\alpha0}|^{2}})=\xi
_{\alpha,\Delta}^{2}|\Delta_{\alpha0}|/\sqrt{\omega^{2}+|\Delta_{\alpha0} |^{2}}$,
and $\xi_{\alpha,\Delta}^{2}\equiv D_{\alpha}/(2|\Delta_{\alpha0}|)$. The
self-consistency condition for corrections can be written as
\end{subequations}
\begin{align}
2\pi T\!\sum_{\omega>0}  &  \left[  \frac{1}{\sqrt{\omega^{2}\!+\!|\Delta
_{\alpha0}|^{2}}}\left(  \tilde{\Phi}_{\alpha}\!-\!\frac{\Delta_{\alpha
0}\operatorname{Re}\left[  \tilde{\Phi}_{\alpha}\Delta_{\alpha0}^{\ast
}\right]  }{\omega^{2}+|\Delta_{\alpha0}|^{2}}\right)  \!-\!\frac
{\tilde{\Delta}_{\alpha}}{\omega}\right] \nonumber\\
&  =\sum_{\beta}w_{\alpha\beta}\tilde{\Delta}_{\beta}-\ln\frac{T_{c}}{T}
\tilde{\Delta}_{\alpha} \label{SelfConPMLin1}
\end{align}
with $w_{\alpha\beta}=\lambda_{\alpha\beta}^{-1}-\lambda^{-1}\delta _{\alpha\beta}$
and $\lambda$ is the largest eigenvalue of the matrix $\lambda_{\alpha\beta}$. The
matrix $w_{\alpha\beta}$ is degenerate, $w_{11}w_{22}-w_{12}w_{21}=0$, and its
components are given by
\begin{equation}
\dbinom{w_{11}}{w_{22}}=\frac{\sqrt{\lambda_{-}^{2}/4\!+\!\lambda_{12}\lambda
_{21}}\!\mp\!\lambda_{-}/2}{\det\lambda},\ w_{12}=\!-\frac{\lambda_{12}}{\det
\lambda} \label{wab}
\end{equation}
with $\lambda_{-}\equiv\lambda_{11}-\lambda_{22}$ and $\det\lambda
\equiv\lambda_{11}\lambda_{22}-\lambda_{12}\lambda_{21}$.

Similar to the $s$-wave case, we can split $\tilde{\Phi}_{\alpha}$ into the
contributions induced by the boundary condition and by the correction to the gap
parameter, $\tilde{\Phi}_{\alpha}=\tilde{\Phi}_{\alpha,b}+\tilde{\Phi
}_{\alpha,\Delta}$. The equation and the boundary condition for $\tilde{\Phi
}_{\alpha,b}(x)$ are
\begin{subequations}
\begin{align}
&  \xi_{\alpha,\omega}^{2}\tilde{\Phi}_{\alpha,b}^{\prime\prime}-\tilde{\Phi
}_{\alpha,b}=0,\label{Eq-Phi-PMb}\\
&  \xi_{\alpha}\tilde{\Phi}_{\alpha,b}^{\prime}=\!-\frac{1}{\gamma_{B\alpha}
}\frac{\sqrt{\omega^{2}\!+\!|\Delta_{\alpha0}|^{2}}}{\sqrt{\omega
^{2}\!+\!\Delta_{s0}^{2}}}\left(  \Delta_{s0}\!-\Delta_{\alpha0}\right)  .
\label{BC-PhiPMb}
\end{align}
The solution for $\tilde{\Phi}_{\alpha,b}(x)$ is given by
\end{subequations}
\begin{align}
\tilde{\Phi}_{\alpha,b}(x)  &  =\frac{\xi_{\alpha,\omega}}{\xi_{\alpha}}
\frac{\sqrt{\omega^{2}+|\Delta_{\alpha0}|^{2}}}{\gamma_{B\alpha}\sqrt
{\omega^{2}+\Delta_{s0}^{2}}}\nonumber\\
\times&  \frac{\cosh\left[  (x-d_{\mathrm{\pm}})/\xi_{\alpha,\omega}\right]
}{\sinh\left(  d_{\mathrm{\pm}}/\xi_{\alpha,\omega}\right)  }(\Delta
_{s0}-\Delta_{\alpha0}). \label{Phi-PMbSol}
\end{align}
The component $\tilde{\Phi}_{\alpha,\Delta}(x)$ has to be found from the following
equation and boundary conditions
\begin{subequations}
\begin{align}
&  \xi_{\alpha,\omega}^{2}\tilde{\Phi}_{\alpha,\Delta}^{\prime\prime}
-\tilde{\Phi}_{\alpha,\Delta}=-\tilde{\Delta}_{\alpha},\label{Eq-Phi-PMD}\\
&  \tilde{\Phi}_{\alpha,\Delta}^{\prime}=0\text{ for }x=0,d_{\mathrm{\pm}}.
\label{BC-PhiPMD}
\end{align}
We can again find $\tilde{\Phi}_{\alpha,\Delta}(x)$ and $\tilde{\Delta
}_{\alpha}(x)$ using Fourier expansions, $\tilde{\Phi}_{\alpha,\Delta
}(x)\!=\!\sum_{m}\tilde{\Phi}_{\alpha,\Delta,m}\cos\left(  q_{m}x\right)  $,
$\tilde{\Delta}_{\alpha}(x)\!=\!\sum_{m}\tilde{\Delta}_{\alpha,m}\cos\left(
q_{m}x\right)  $ with $q_{m}\!=\!m\pi/d_{\mathrm{\pm}}$. From Eq. (\ref{Eq-Phi-PMD})
we immediately find
\end{subequations}
\begin{equation}
\tilde{\Phi}_{\alpha,\Delta,m}=\frac{\tilde{\Delta}_{\alpha,m}}{1+\xi
_{\alpha,\omega}^{2}q_{m}^{2}}. \label{PhiPMDm-Dm}
\end{equation}

As follows from the structure of the the self-consistency equation
(\ref{SelfConPMLin1}), the responses of the order parameters $\tilde{\Delta
}_{\alpha}$ are different in the amplitude and phase channels. To proceed, we split
all quantities into the amplitude and phase components, as illustrated in Fig.
\ref{Fig-DeltasFrustrState}, $X=X^{a}+X^{\phi}$,
\[
X^{a}=\frac{\Delta_{\alpha0}\operatorname{Re}[X\Delta_{\alpha0}^{\ast}
]}{|\Delta_{\alpha0}|^{2}};\ X^{\phi}=X-\frac{\Delta_{\alpha0}
\operatorname{Re}[X\Delta_{\alpha0}^{\ast}]}{|\Delta_{\alpha0}|^{2}},
\]
where $X$ stands for $\tilde{\Delta}_{\alpha}$, $\tilde{\Phi}_{\alpha,b}$,
$\tilde{\Phi}_{\alpha,\Delta}$, and $\Delta_{s0}$. Such decomposition of
$\Delta_{s0}$ is illustrated in Fig. \ref{Fig-DeltasFrustrState}. Explicitly, we can
write, $\Delta_{s0}^{a}=(\Delta_{10}/|\Delta_{10}|)|\Delta_{s0} |\cos\phi$ and
$\Delta_{s0}^{\phi}=-i(\Delta_{10}/|\Delta_{10}|)|\Delta _{s0}|\sin\phi$.
Substituting $\tilde{\Phi}_{\alpha,\Delta}$ into the self-consistency equation
(\ref{SelfConPMLin1}), we obtain the following equations for
$\tilde{\Delta}_{\alpha,m}^{a}$ and $\tilde{\Delta}_{\alpha ,m}^{\phi}$
\begin{subequations}
\label{EqDpm}
\begin{align}
&  \sum_{\beta}\left(  w_{\alpha\beta}\!-\!\Sigma_{\alpha,m}^{a}\delta
_{\alpha\beta}\right)  \tilde{\Delta}_{\beta,m}^{a}\!=\!2\pi T\sum_{\omega
>0}\!\frac{\omega^{2}\tilde{\Phi}_{\alpha,b,m}^{a}}{\left(  \omega
^{2}\!+\!|\Delta_{\alpha0}|^{2}\right)  ^{3/2}},\label{EqDPMa}\\
&  \Sigma_{\alpha,m}^{a}\!=\!2\pi T\!\sum_{\omega>0}\!\left[ \frac{\omega
^{2}}{\left( \omega^{2}\!+\!|\Delta_{\alpha0}|^{2}\right)^{3/2}\!
\left(1\!+\!\xi_{\alpha,\omega}^{2}q_{m}^{2}\right)  }\!-\!\frac{1}{\omega}\right]
\!+\!\ln\!\frac{T_{c}}{T},\nonumber\\
&  \sum_{\beta}\left(  w_{\alpha\beta}\!-\!\Sigma_{\alpha,m}^{\phi}
\delta_{\alpha\beta}\right)  \tilde{\Delta}_{\beta,m}^{\phi}\!=\!2\pi
T\sum_{\omega>0}\!\frac{\tilde{\Phi}_{\alpha,b,m}^{\phi}}{\sqrt{\omega
^{2}\!+\!|\Delta_{\alpha0}|^{2}}},\label{EqDPMphi}\\
&  \Sigma_{\alpha,m}^{\phi}\!=\!2\pi T\sum_{\omega>0}\left[  \frac{1}
{\sqrt{\omega^{2}\!+\!|\Delta_{\alpha0}|^{2}}}\frac{1}{1\!+\!\xi
_{\alpha,\omega}^{2}q_{m}^{2}}\!-\!\frac{1}{\omega}\right]  \!+\!\ln
\frac{T_{c}}{T},\nonumber
\end{align}
where the Fourier components $\tilde{\Phi}_{\alpha,b,m}^{a}$ and $\tilde{\Phi
}_{\alpha,b,m}^{\phi}$ can be computed explicitly from Eq.\ (\ref{Phi-PMbSol} ),
\end{subequations}
\begin{align}
\binom{\tilde{\Phi}_{\alpha,b,m}^{a}}{\tilde{\Phi}_{\alpha,b,m}^{\phi}}  &
=\frac{\left(  2-\delta_{m}\right)  }{\gamma_{B\alpha}}\frac{\xi
_{\alpha,\omega}^{2}/\left(  d_{\mathrm{\pm}}\xi_{\alpha}\right)  }
{1+\xi_{\alpha,\omega}^{2}q_{m}^{2}}\nonumber\\
&  \times\frac{\sqrt{\omega^{2}+|\Delta_{\alpha0}|^{2}}}{\sqrt{\omega
^{2}+\Delta_{s0}^{2}}}\binom{\Delta_{s0}^{a}\!-\!\Delta_{\alpha0}}{\Delta
_{s0}^{\phi}}. \label{PhiPMbm}
\end{align}
Solutions of Eqs. (\ref{EqDpm}) are
\begin{subequations}
\label{DPMSol}
\begin{align}
\tilde{\Delta}_{\alpha,m}^{a}  &  =2\pi T\sum_{\beta,\omega>0}U_{m,\alpha
\beta}^{a}\frac{\omega^{2}\tilde{\Phi}_{\beta,b,m}^{a}}{\left(  \omega
^{2}+|\Delta_{\beta0}|^{2}\right)  ^{3/2}},\label{DPMaSol}\\
\tilde{\Delta}_{\alpha,m}^{\phi}  &  =2\pi T\sum_{\beta,\omega>0}
U_{m,\alpha\beta}^{\phi}\frac{\tilde{\Phi}_{\beta,b,m}^{\phi}}{\sqrt
{\omega^{2}+|\Delta_{\beta0}|^{2}}}, \label{DPMphiSol}
\end{align}
where the matrices $U_{m,\alpha\beta}^{a,\phi}=\left[  w_{\alpha\beta}
-\Sigma_{\alpha,m}^{a,\phi}\delta_{\alpha\beta}\right]  ^{-1}$ in the two-band case
are given by
\end{subequations}
\begin{align}
U_{m,\alpha\beta}^{a,\phi}  &  =\frac{1}{D_{U,m}^{a,\phi}}
\begin{bmatrix}
w_{22}\!-\!\Sigma_{2,m}^{a,\phi} & -w_{12}\\
-w_{21} & w_{11}\!-\!\Sigma_{1,m}^{a,\phi}
\end{bmatrix}
,\label{Uaphiab}\\
D_{U,m}^{a,\phi}  &  =-\Sigma_{2,m}^{a,\phi}w_{11}\!-\!\Sigma_{1,m}^{a,\phi
}w_{22}+\Sigma_{1,m}^{a,\phi}\Sigma_{2,m}^{a,\phi}.\nonumber
\end{align}

The equation for the phase component at $m=0$ requires special attention. We note
that the equation for the bulk gaps have the form
\[
\sum_{\beta}\left(  w_{\alpha\beta}-\Sigma_{\alpha,0}^{\phi}\delta
_{\alpha\beta}\right)  \Delta_{\beta0}=0
\]
meaning that Eq. (\ref{EqDPMphi})\ at $m=0$ is actually degenerate, i.e., its
determinant vanishes, $D_{U,0}^{a,\phi}=$ $-w_{11}\Sigma_{2,0}^{\phi}
-w_{2}\Sigma_{1,0}^{\phi}+\Sigma_{1,0}^{\phi}\Sigma_{2,0}^{\phi}=0$. This degeneracy
reflects gauge invariance with respect to identical phase change of the order
parameters $\Delta_{\alpha0}$. This means that the equation for
$\tilde{\Delta}_{\alpha,0}^{\phi}$ only has solution if its right-hand side
satisfies certain condition which, using the bulk gap ratio $\Delta
_{10}/\Delta_{20}=-w_{12}/(w_{11}-\Sigma_{1,0}^{\phi})=-(w_{22}-\Sigma
_{2,0}^{\phi})/w_{21}$, can be written as
\begin{equation}
2\pi T\sum_{\omega>0}\left(  \frac{w_{21}\Delta_{10}\tilde{\Phi}_{1,b,0}
^{\phi}}{\sqrt{\omega^{2}\!+\!|\Delta_{10}|^{2}}}+\frac{w_{12}\Delta
_{20}\tilde{\Phi}_{2,b,0}^{\phi}}{\sqrt{\omega^{2}\!+\!|\Delta_{20}|^{2}}
}\right)  =0. \label{CondPM0}
\end{equation}
The meaning of this condition is that the total \textquotedblleft
torque\textquotedblright\ from the interface forcing unform phase rotation of the
$s_{\pm}$ gap parameters has to vanish. Using the relation $w_{21}
/w_{12}=\lambda_{21}/\lambda_{12}=\nu_{1}/\nu_{2}$ and Eq. (\ref{PhiPMbm}) we can
rewrite this condition as
\[
2\pi T\sum_{\alpha,\omega>0}\frac{\xi_{\alpha,\omega}^{2}}{\xi_{\alpha}}
\frac{\nu_{\alpha}\Delta_{\alpha0}}{\sqrt{\omega^{2}+\Delta_{s0}^{2}}}
\frac{\Delta_{s0}^{\phi}}{\gamma_{B\alpha}}=0.
\]
with $\Delta_{s0}^{\phi}=\Delta_{s0}\sin\phi$. Moreover, for the combination
$\xi_{\alpha,\omega}^{2}\nu_{\alpha}/\xi_{\alpha}$ we obtain
\[
\frac{\xi_{\alpha,\omega}^{2}\nu_{\alpha}}{\xi_{\alpha}}\propto\frac
{D_{\alpha}\nu_{\alpha}}{\xi_{\alpha}\sqrt{\omega^{2}+|\Delta_{\alpha0}|^{2}}
}\propto\frac{1}{\gamma_{\alpha}\sqrt{\omega^{2}+|\Delta_{\alpha0}|^{2}}},
\]
which allows us to present the condition in the form
\[
2\pi T\sum_{\alpha,\omega>0}\frac{1}{\tilde{\gamma}_{B\alpha}}\frac
{\Delta_{\alpha0}\sin\phi}{\sqrt{\omega^{2}+|\Delta_{\alpha0}|^{2}}
\sqrt{\omega^{2}+\Delta_{s0}^{2}}}=0.
\]
We immediately recognize that this condition is equivalent to Eq.\ (\ref{ImCond})
(vanishing of the total Josephson current flowing through the interface). With this
condition, Eq.\ (\ref{PhiPMbm}) for $\tilde{\Delta }_{\alpha,0}^{\phi}$ determines
interface-induced phase shifts $\varphi _{\alpha}\ll1$ of the averaged order
parameters with respect to zero-order phases $\phi$ and $\phi-\pi$, see Fig.\
\ref{Fig-DeltasFrustrState}. These phase shifts are defined by relation
$\tilde{\Delta}_{\alpha,0}^{\phi }=i\varphi_{\alpha}\Delta_{\alpha0}$. The same
phase shifts appear in the phenomenological frustrated Josephson junction model. As
the average phase shift $(\varphi_{1}+\varphi_{2})/2$ can be absorbed into $\phi$,
we can set it zero and take $\varphi_{1,2}=\pm\varphi/2$. From Eq. (\ref{ImCond}),
taking into account the above condition, we derive
\begin{equation}
\varphi\!=\!-\frac{2\pi T}{d_{\mathrm{\pm}}w_{12}|\Delta_{20}|}\sum_{\omega
>0}\frac{\xi_{1,\omega}^{2}/\xi_{1}}{\sqrt{\omega^{2}\!+\!\Delta_{s0}^{2}}
}\frac{|\Delta_{s0}|}{\gamma_{B1}}\sin\phi\label{dPhi0}
\end{equation}
One can verify that this result does not change with the switching of indices
$1\leftrightarrow2$ in the right hand side. Using the expression for the partial
Josephson energy, Eq.\ (\ref{PartJosEner}), this result can be rewritten as
\begin{equation}
\varphi=-\frac{E_{J,1}}{d_{\mathrm{\pm}}\nu_{1}w_{12}|\Delta_{20}||\Delta
_{10}|}\sin\phi. \label{dTheta}
\end{equation}
For weak interband coupling, $|\lambda_{12}|,|\lambda_{21}|\ll\lambda
_{11},\lambda_{22}$, the parameter $\mathcal{E}_{12}=\nu_{1}w_{12}|\Delta
_{20}||\Delta_{10}|$ represents the interband energy and this result coincides with
the result obtained within the frustrated Josephson junction model in the case
$E_{J,1}\ll d_{\mathrm{\pm}}\mathcal{E}_{12}$. In this situation $w_{12}>0$ and
$\varphi<0$. However, in contrast to this model, in a general situation the phases
$\varphi_{\alpha}$ do not fully determine the energy of the $s_{\pm}$
superconductor. Moreover, it was argued that for the iron-based superconductors the
pairing is dominated by the interband coupling, i.e., the opposite inequality holds,
$|\lambda_{12}|,|\lambda_{21}|\gg\lambda _{11},\lambda_{22}$. In this case, as
$\lambda_{12}<0$ and $\lambda _{11}\lambda_{22}-\lambda_{12}\lambda_{21}<0$, we have
$w_{12}<0$ meaning that $\varphi>0$. Fig. \ref{Fig-DeltasFrustrState} actually
illustrates this situation. Using the results presented in this Appendix, we derive
in the main text the amplitude of the $\sin2\phi$ term in the Josephson current
which determines the width of the TRSB state.


\begin{thebibliography}{99}

\bibitem{Reviews}D. C. Johnston, Adv. Phys. \textbf{59}, 803 (2010); J-P.
Paglione and R. L. Green, Nature Phys. \textbf{6}, 645 (2010);P. C. Canfield and S.
L. Bud'ko, Annu. Rev. Cond. Mat. Phys. \textbf{1}, 27 (2010).

\bibitem{ReviewsTheory}P. J. Hirschfeld, M. M. Korshunov, and I. I. Mazin,
Rep. Prog. Phys., \textbf{74}, 124508 (2011); A. V. Chubukov, Annu. Rev. Cond. Mat.
Phys. \textbf{3}, 57 (2012).

\bibitem{BoeriPRL08}L. Boeri, O. V. Dolgov, and A. A. Golubov, Phys. Rev.
Lett. \textbf{101}, 026403 (2008); A. Subedi, L. Zhang, D. J. Singh, and M. H. Du,
Phys. Rev. B \textbf{78}, 134514 (2008); T. Yildirim, Phys. Rev. Lett. \textbf{102},
037003 (2009).

\bibitem{MazinPRL08}I. I. Mazin, D. J. Singh, M. D. Johannes, and M. H. Du,
Phys. Rev. Lett. \textbf{101}, 057003 (2008).

\bibitem{KurokiPRL08}K. Kuroki, S. Onari, R. Arita, H. Usui, Y. Tanaka, H.
Kontani, and H. Aoki, Phys. Rev. Lett. \textbf{101}, 087004 (2008).

\bibitem{SeoPRL08}K. Seo, B. A. Bernevig, and J. Hu, Phys. Rev. Lett.
\textbf{101}, 206404 (2008).

\bibitem{GraserNJP09}S. Graser, T. A. Maier, P. J. Hirschfeld, and D. J.
Scalapino, New J. Phys. \textbf{11}, 025016 (2009).

\bibitem{CvetkovicEPL09}V. Cvetkovic and Z. Tesanovic, EPL \textbf{85}, 37002 (2009).

\bibitem{MagRes}A. D. Christianson, E. A. Goremychkin, R. Osborn, S.
Rosenkranz, M. D. Lumsden, C. D. Malliakas, l. S. Todorov, H. Claus, D. Y. Chung, M.
G. Kanatzidis, R. I. Bewley, and T. Guidi, Nature \textbf{456}, 930 (2008); M.D.
Lumsden, A.D. Christianson J. Phys.: Condens. Matter, \textbf{22}, 203203 (2010).

\bibitem{LaplacePRB09}Y. Laplace, J. Bobroff, F. Rullier-Albenque, D. Colson, and A. Forget,
Phys. Rev. B \textbf{80}, 140501(R) (2009).

\bibitem{Fernandes10}R. M. Fernandes, D. K. Pratt, W. Tian, J. Zarestky, A. Kreyssig, S. Nandi,
Min Gyu Kim, A. Thaler, Ni Ni, P. C. Canfield, R. J. McQueeney, J. Schmalian, and A.
I. Goldman Phys. Rev. B, \textbf{81}, 140501(R) (2010).

\bibitem{CoexTheory}A. B. Vorontsov, M. G. Vavilov, and A. V. Chubukov, Phys. Rev. B \textbf{81}, 174538(2010);
R. M. Fernandes, J. Schmalian, Phys. Rev. B \textbf{82}, 014521(2010).

\bibitem{HanaguriSci10}T. Hanaguri, S. Niitaka, K. Kuroki, H. Takagi, Science \textbf{328}, 474 (2010).

\bibitem{K122}J. Guo, S. Jin, G. Wang, S. Wang, K Zhu, T. Zhou, M. He, and X.
Chen, Phys. Rev. B \textbf{82}, 180520 (2010).

\bibitem{AgterbergPRB02}D. F. Agterberg, E. Demler, and B. Janko,
Phys. Rev. B \textbf{66}, 214507 (2002).

\bibitem{Ng}T. K. Ng and N. Nagaosa, Europhys. Lett. \textbf{87}, 17003 (2009).

\bibitem{Linder}J. Linder, I. B. Sperstad, A. Sudbo, Phys. Rev. B
\textbf{80}, 020503(R) (2009).

\bibitem{SperstadPRB09}I. B. Sperstad, J. Linder, A. Sudbo, Phys. Rev. B
\textbf{80}, 144507 (2009).

\bibitem{VS}V. Stanev and Z. Te\v sanovi\' c, Phys. Rev. B \textbf{81},
134522 (2010).

\bibitem{Ota}Y. Ota, M. Machida, T. Koyama, and H. Matsumoto, Phys. Rev. Lett.
\textbf{102}, 237003 (2009); Y. Ota, M. Machida, and T. Koyama, Phys. Rev. B
\textbf{82}, 140509(R) (2010); Y. Ota, M. Machida and T. Koyama, Phys. Rev. B
\textbf{83}, 060503(R) (2011).

\bibitem{Berg}E. Berg, N. H. Lindner, and T. Pereg-Barnea, Phys. Rev. Lett.
\textbf{106}, 147003 (2011).

\bibitem{ProxFingeprintEPL11}A.E. Koshelev and V. Stanev, Europhys.\ Lett.
\textbf{96}, 27014 (2011); V. Stanev and A.E. Koshelev,  arXiv:1207.5565.

\bibitem{Lin}S. Z. Lin, Phys. Rev. B \textbf{86}, 014510 (2012).

\bibitem{Levchenko}S. Apostolov and A. Levchenko, arXiv:1210.1875.

\bibitem{SeidelSUST11} P. Seidel, Supercond. Sci. Technol. \textbf{24}, 043001
(2011).

\bibitem{BuzdinReview}A. I. Buzdin, Rev. Mod. Phys. \textbf{77}, 935 (2005).

\bibitem{fi-junction}A. Buzdin and A. E. Koshelev, Phys. Rev. B \textbf{67}, 220504(R) (2003);
H. Sickinger, A. Lipman, M. Weides, R. G. Mints, H. Kohlstedt, D. Koelle, R.
Kleiner, and E. Goldobin,  Phys. Rev. Lett. \textbf{109}, 107002 (2012).

\bibitem{Usadel}K. Usadel, Phys. Rev. Lett. \textbf{25}, 560 (1970).

\bibitem{KL}M. Yu. Kupriyanov and V. F. Lukichev, Zh. Eksp. Teor. Fiz. \textbf{94},
139 (1988) [Sov. Phys. JETP \textbf{67}, 1163 (1988)].

\bibitem{Golubov1}A. A. Golubov, E. P. Houwman, J. G. Gijsbertsen, V. M.
Krasnov, J. Flokstra, H. Rogalla, and M. Yu. Kupriyanov, Phys. Rev. B \textbf{51},
1073 (1995).

\bibitem{Golubov2}A. Brinkman, A.A. Golubov, M.Yu. Kupriyanov, Phys. Rev. B,
\textbf{69}, 214407 (2004).

\bibitem{JosCurrTxtbk}V. Ambegaokar and A. Baratoff, Phys. Rev. Lett.
\textbf{10}, 486 (1963); A. Barone and G. Patern\`{o}, \textit{Physics and
Applications of the Josephson Effect}, A Wiley-Interscience Publication, John Wiley
\& Sons, New York/ Chichester/Brisbane/Toronto/Singapore, 1982, p.52.

\bibitem{GolubovKuprPZHETF05}A.A. Golubov and M. Yu. Kupriyanov, Pis'ma
Zh. Eksp. Teor. Fiz., \textbf{81}, 419 (2005) [JETP Lett. \textbf{81}, 335 (2005)].

\bibitem{Note39Bulk} Even though, in general, the right hand side of Eq.\
(\ref{TRSBwidth}) depends on the ratio of the boundary resistances $R_{B1}/R_{B2}$,
in the transition region this ratio is approximately fixed by the condition
$E_{J1}=E_{J2}$ and, as consequence, also mostly depends on bulk parameters.

\bibitem{RyazSFS0Pi}V. V. Ryazanov, V. A. Oboznov, A. Yu. Rusanov, A. V.
Veretennikov, A. A. Golubov, and J. Aarts, Phys. Rev. Lett. \textbf{86}, 2427
(2001).

\bibitem{GapAnis}K. Umezawa, Y. Li, H. Miao, K. Nakayama, Z.-H. Liu, P. Richard, T. Sato, J. B. He, D.-M. Wang,
G. F. Chen, H. Ding, T. Takahashi, and S.-C. Wang, Phys. Rev. Lett. \textbf{108},
037002 (2012); M. P. Allan, A. W. Rost, A. P. Mackenzie, Yang Xie, J. C. Davis, K.
Kihou, C. H. Lee, A. Iyo, H. Eisaki, T.-M. Chuang, Science \textbf{336}, 563 (2012).

\bibitem{nodes}J. D. Fletcher, A. Serafin, L. Malone, J. G. Analytis, J. H. Chu, A. S. Erickson, I.
R. Fisher, and A. Carrington, Phys. Rev. Lett. \textbf{102}, 147001 (2009); K.
Hashimoto, M. Yamashita, S. Kasahara, Y. Senshu, N. Nakata, S. Tonegawa, K. Ikada,
A. Serafin, A. Carrington, T. Terashima, H. Ikeda, T. Shibauchi, and Y. Matsuda,
Phys. Rev. B \textbf{81}, 220501(R) (2010).

\bibitem{IsGaps}P. Richard and T. Sato and K. Nakayama and T. Takahashi and H. Ding,
Rep. Prog. Phys., \textbf{74} 124512 (2011).
\end{thebibliography}
\end{document}